\pdfoutput=1 % ensure pdflatex
\documentclass[10pt, twoside]{predoc2}

\usepackage[sans]{distoperators}

\usepackage{graphicx}

\usepackage{amsfonts}
\usepackage{amsbsy}
\usepackage{amsmath}
\usepackage{mathrsfs}
\usepackage{amsthm}

\usepackage{float}
\usepackage{subcaption}
\usepackage{caption}
\usepackage{booktabs}
\usepackage{tabularx}

\usepackage{IEEEtrantools}

\newtheoremstyle{mystyle}%                % Name
  {}%                                     % Space above
  {}%                                     % Space below
  {\itshape}%                             % Body font
  {}%                                     % Indent amount
  {\bfseries\sffamily}%                   % Theorem head font
  {.}%                                    % Punctuation after theorem head
  { }%                                    % Space after theorem head, ' ', or \newline
  {}%                                     % Theorem head spec (can be left empty, meaning `normal')
\theoremstyle{mystyle}

\usepackage{lipsum}

\title{Predictive Assessment and Comparison of Bayesian Survival Models for Cancer Recurrence}
\author[1,2]{Saku Suorsa}
\author[1,2]{Aki Vehtari}
\affil[1]{ELLIS Institute Finland}
\affil[2]{Department of Computer Science, Aalto University, Finland}
\date{}

\makeatletter
\def\runauthor{Suorsa and Vehtari}
\def\runtitle{Assessing Bayesian survival models}
\makeatother

\begin{document}
\maketitle
\thispagestyle{empty}

\begin{abstract}
Complex data features, such as unmodelled censored event times and variables with
time-dependent effects, are common in cancer recurrence studies and pose challenges
for Bayesian survival modelling. Current methodologies for predictive
model checking and comparison often fail to adequately address these features. This
paper bridges that gap by introducing new, targeted recommendations for predictive
assessment and comparison of Bayesian survival models. Our
recommendations cover a variety of different scenarios and models. Accompanying
code together with our implementations to open source software help in replicating
the results and applying our recommendations in practice.
\end{abstract}


\section{Introduction}

The motivation for this work is to improve the predictive model checking and the predictive model comparison of Bayesian survival models with the aim of analysing the gastrointestinal stromal tumour (GIST) data set that was collected over multiple decades in a collaborative effort of multiple research centres around the world. The data set has interesting aspects related to predictive model checking and predictive model comparison that are common in survival analysis but have not received much attention in the Bayesian literature and software implementations. A large portion of the event times in the GIST data set are censored, but the independent censoring process is not known nor modelled. This means that the full generative model is not available, which complicates predictive model checking and predictive model comparison. The GIST data set also has variables with clear time-dependent effects, which encourages the building of more complex models that further complicate predictive model comparison.

All GIST-related analysis presented in this paper is performed on simulated data, since we are prohibited from publishing results from the real data set. In our other examples, we use simulated data or the Rotterdam breast cancer data set \citep{royston2013}. All the code and data used in this paper are freely available. \footnote{Code and data required to replicate the results in this paper are freely available at \href{https://github.com/Sakuski/Bayesian-Cancer-Recurrence-Prediction}{https://github.com/Sakuski/Bayesian-Cancer-Recurrence-Prediction}.}

\subsection{Our contributions}

We present new recommendations for predictive model checking and predictive model comparison in Bayesian survival modelling. We give recommendations on how to perform predictive checking for different survival models in case there are no censored event times in the data, and in case there are censored event times in the data and the censoring process is not modelled. We also give recommendations for performing predictive model comparison when some of the models cause problems in the regular Pareto smoothed importance sampling leave-one-out cross-validation (PSIS-LOO CV) based comparison. The recommendations are summarised in Section \ref{sec:recommendations} and justified in Section \ref{sec:predictive_model_checking} and Section \ref{sec:predictive_model_comparison}. Some of our recommendations were included as software implementations in the release v.1.13.0 of bayesplot R package \citep{bayesplot}.

\subsection{Relation to previous work}

Although the roots of survival analysis go back to the 17th century \citep{graunt1662}, it is the ground-breaking work by \cite{kaplan1958}, \cite{cox1972}, \cite{nelson1972}, and \cite{aalen1978} in the later part of the 20th century that leads to the modern framework. In the early years, survival analysis mainly followed the frequentist approach, but as computational capabilities grew larger, the Bayesian approach became a common alternative. Notable early contributions in this area include the work by \cite{kalbfleisch1978} and \cite{ibrahim2001}.

In recent years, the concept of Bayesian workflow \citep[e.g., ][]{gelman2020_workflow} as a structural process has been established to help with practical aspects of Bayesian analysis. Different parts of the Bayesian workflow, including inference, predictive model checking, and predictive model comparison, have been automated in software packages, such as brms \citep{brms2017, brms2018, brms2021}, bayesplot \citep{bayesplot, gabry2019}, and loo \citep{vehtari2024_loo}, respectively.

Bayesian methodology has been applied in previous work related to gastrointestinal stromal tumours. For example, \cite{joensuu2012} utilised a modified Cox proportional hazards model fitted with Bayesian inference to non-linearly model recurrence free survival. In addition, \cite{joensuu2014} modelled the hazard of recurrence of GIST with Bayesian methods to optimise the timing of CT scans.

The relationship of our work and previous research can be summarised in the following words. We take a close look at two important steps in the established Bayesian workflow: predictive model checking and predictive model comparison. We improve these steps in the context of survival modelling through our contributions. We also improve and give improvement suggestions to the relevant software tools in this area. In addition, we continue to apply Bayesian survival modelling in the analysis of simulated data on recurrence of gastrointestinal stromal tumours.

\subsection{Paper outline}

The remainder of this paper is organised as follows. We briefly cover the models of our focus in Section \ref{sec:survival_models}. In Sections \ref{sec:predictive_model_checking} and \ref{sec:predictive_model_comparison}, we present our contributions to the Bayesian workflow for survival modelling: improvements to predictive model checks and new suggestions for predictive model comparison. In Section \ref{sec:case_study}, we demonstrate the methodology covered in earlier sections of the paper in a case study on a data set related to gastrointestinal stromal tumours. Finally, we summarise our contributions in the form of recommendations in Section \ref{sec:recommendations} and discuss limitations and potential directions for future research in Section \ref{sec:discussion}.

\section{Survival models}
\label{sec:survival_models}

In this paper, we focus only on models that are supported by the brms R package \citep{brms2017, brms2018, brms2021}. The brms framework allows for the inclusion of censoring and left-truncation in survival modelling. The possible model types include the common proportional hazards (PH) and accelerated failure time (AFT) models, but also a Bernoulli model that requires a bit more manual crafting.

The Bernoulli model, which we introduce here, uses discrete time. Thus, in order to apply the model in a practical cancer recurrence setting, one must first discretise the continuous follow-up time. This involves choosing a suitable interval length (e.g., one month or one year) and dividing each subject's observation period into these discrete time intervals.

For each individual and for each discrete time interval in which they are observed and at risk, we model the probability of cancer recurrence using a Bernoulli model. This means that within any given interval, the outcome for a subject is binary: either a recurrence occurs, or it does not. A key point is that time intervals after a subject has experienced the event (recurrence) or has been censored (e.g., died from other causes, lost to follow-up) are excluded from the analysis.

The Bernoulli model assumes that the probability of recurrence occurring at some point in time for some individual is $\theta$, which depends on the possibly time-dependent covariates. For example, if we are applying a logit link, the relationship between $\theta$ and the covariates is

\begin{equation}
\mathrm{logit}(\theta) = \log \left( \frac{\theta}{1-\theta} \right) = \beta_0 + \beta_1 X_1 + \dots + \beta_p X_p,
\end{equation}
where $X_j$ is the covariate $j$ and $\beta_j$ is its regression coefficient.

This modelling approach requires the data set to be transferred to a long format. Instead of having one row per subject, we create a data point for each discrete time interval for each individual. In this format, a single subject is represented by multiple rows, one for each time interval in which they were observed and at risk. Time-invariant covariates remain constant across all rows for the same subject. The main advantage of the long data format is its ability to easily incorporate time-dependent covariates.

\begin{figure}[H]
  \begin{center}
    \includegraphics[width=0.49\textwidth]{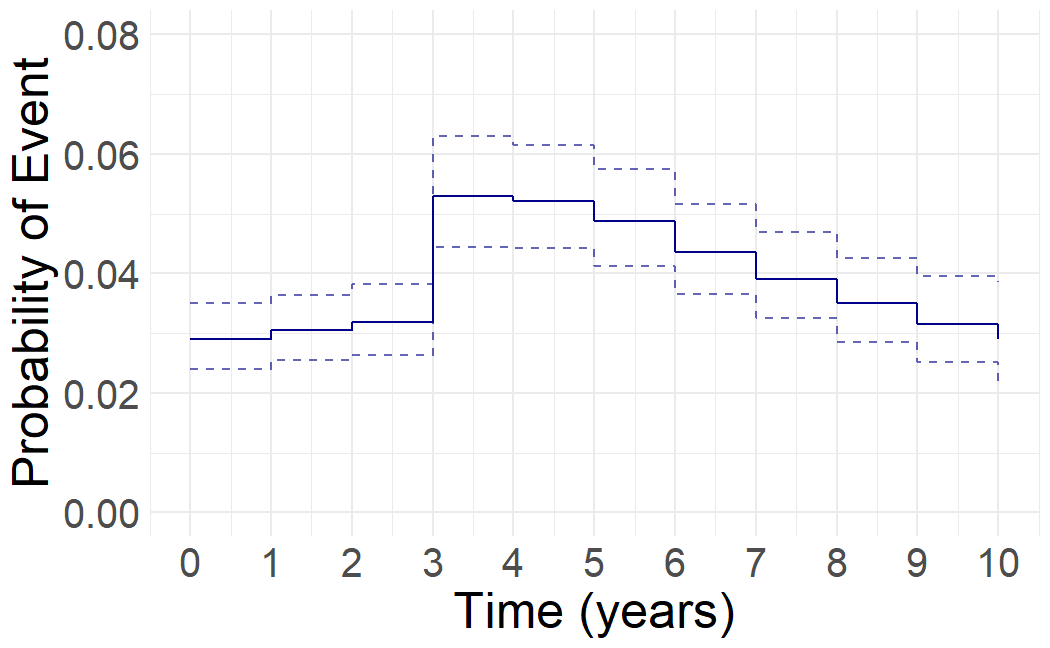}
    \caption{An example of a Bernoulli model applied in cancer recurrence prediction. The model predicts the probability of recurrence for a patient for each year after surgery conditional on the fact that it has not happened in the previous years. The model allows the inclusion of variables with time-dependent effects. The patient in the example receives adjuvant treatment for three years after surgery, and the time-dependent effect of treatment on the predicted probability can be clearly seen in the figure. The figure shows that the predicted probability of event jumps high immediately after the adjuvant treatment ends.}
  \end{center}
\end{figure}

\section{Predictive checking of survival models}
\label{sec:predictive_model_checking}

If the generative model is available, it can be used to perform predictive model checks. Predictive model checking is an important part of the Bayesian workflow. Before fitting the model, it is useful to simulate data from the prior predictive distribution to check that the priors combined with the likelihood produce reasonable data. This is called prior predictive checking \citep{gelman2013}. Similarly, after fitting the model, it is important to verify that the observed data are plausible under the fitted model. This is referred to as posterior predictive checking \citep{gelman2013}. Posterior predictive checking can detect model misspecification and prevent problematic scenarios in leave-one-out cross-validation \citep{Sivula-Magnusson-Vehtari:2025}. Predictive model checks can be performed visually \citep{gabry2019}, which is the focus of this section.

The recommended predictive model checks depend on whether there are censored event times in the data and whether the model to be assessed is parametric, semi-parametric, or the Bernoulli model. However, at the time of writing, brms does not support plotting predictions for the Cox PH model, which is the only semi-parametric model discussed in this paper. Therefore, the different cases that we review in this section are 1. parametric survival models in the absence of censored event times, 2. parametric survival models in the presence of censored event times, and 3. the Bernoulli model.

\subsection{Predictive checking of parametric survival models in the absence of censored event times}

\begin{figure}[t]
  \begin{center}
    \includegraphics[width=0.6\textwidth]{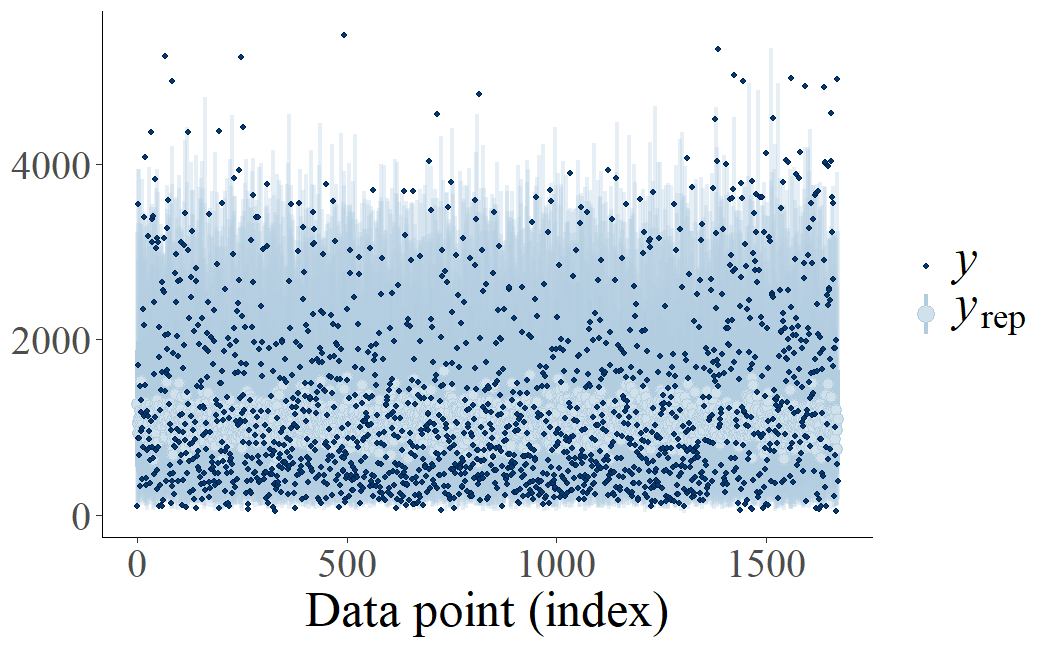}
    \caption{The intervals plot displays observations with dark points, medians of predictive draws with lighter points, and central intervals for predictive draws with vertical bars. When the model fits well, most observations should fall within the corresponding predictive intervals.}
    \label{fig:ppc_intervals_observed}
  \end{center}
\end{figure}

\begin{figure}[t]
  \begin{center}
    \includegraphics[width=0.6\textwidth]{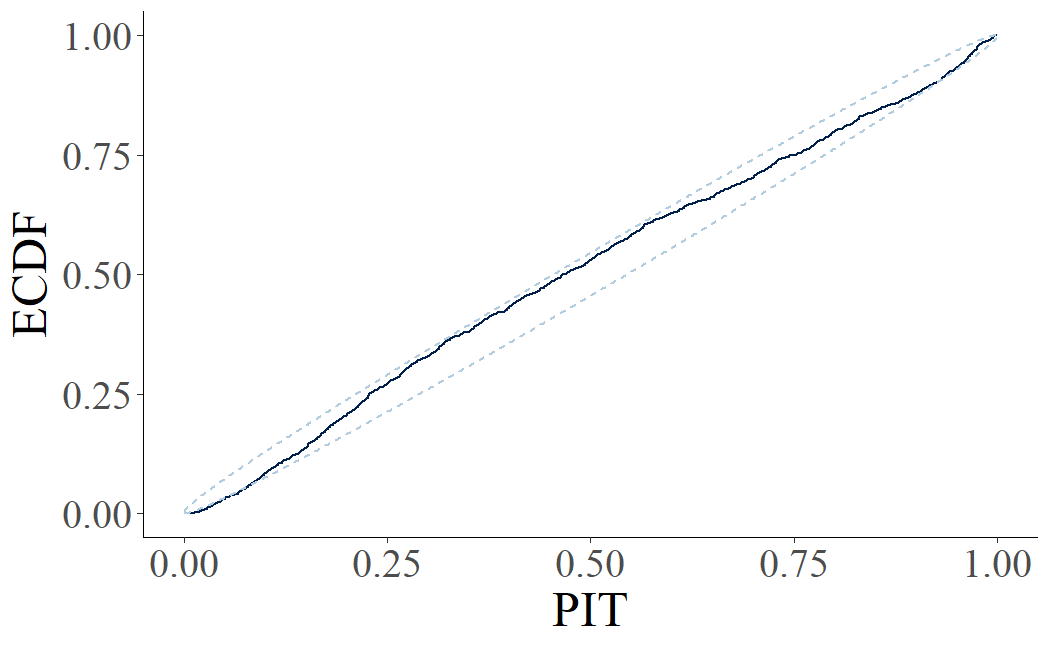}
    \caption{The PIT-ECDF plot visualises the empirical cumulative distribution function (ECDF) of the probability integral transformation (PIT) of observations with respect
to the corresponding predictive draws. The plot also displays central simultaneous
confidence bands that can be used to assess whether observations and predictive
draws come from the same distribution. PIT-ECDF that exceeds the confidence bands indicates a lack of fit.}
    \label{fig:ppc_pit_ecdf_observed}
  \end{center}
\end{figure}

\begin{figure}[t]
  \begin{center}
    \includegraphics[width=0.6\textwidth]{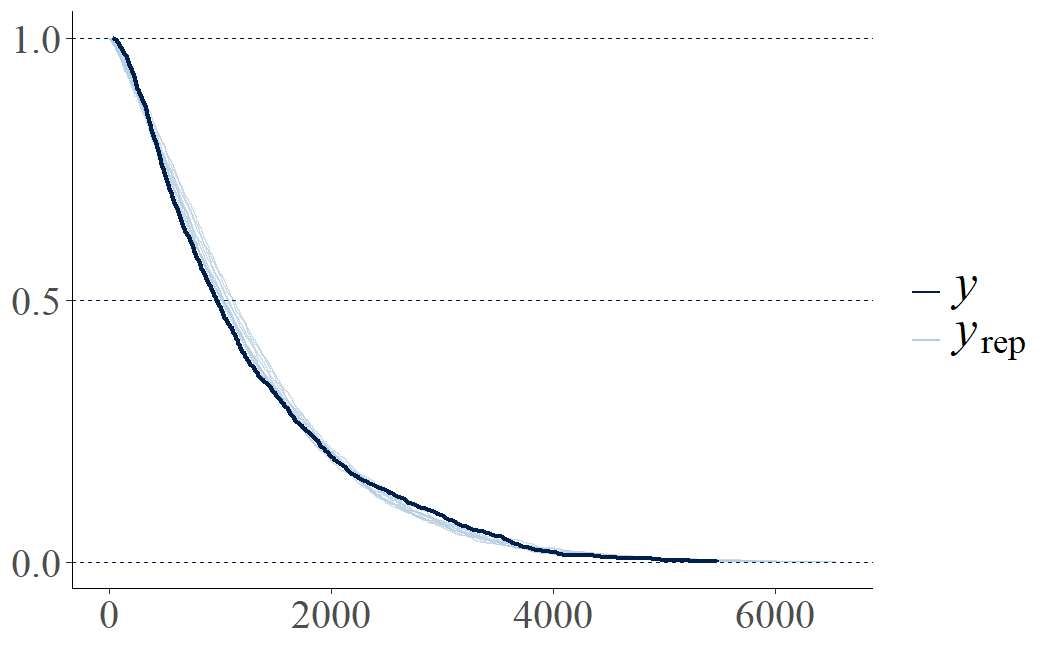}
    \caption{The Kaplan-Meier overlay plot shows the Kaplan-Meier survival curve \citep{kaplan1958} of the observations with a dark colour and the empirical CCDF estimates of the predictive draws with a lighter colour. If the survival curve deviates from the empirical CCDF estimates of the predictive draws, there is probably some lack of fit.}
    \label{fig:ppc_km_overlay_observed}
  \end{center}
\end{figure}

When there are no censored event times in the data set, the full generative model is generally available, which makes the predictive model assessment straightforward. In this case, many diagnostic plots in the brms package can be used directly to graphically assess the predictions. Figures \ref{fig:ppc_intervals_observed}, \ref{fig:ppc_pit_ecdf_observed}, and \ref{fig:ppc_km_overlay_observed} show three common visualisations for assessing the predictions of parametric survival models when the full generative model is available. All three visualisations are for the same data set.

Figure \ref{fig:ppc_intervals_observed} shows a conceptually simple intervals plot. It is a plot that compares the observations with the corresponding central intervals for predictive draws. In a well-fitting model, most observations should fall within these intervals.

Figure \ref{fig:ppc_pit_ecdf_observed} shows a PIT-ECDF plot, which is based on the following mathematical result: Assume a continuous random variable $X$ for which the cumulative distribution function (CDF) is $F_X$. Then, the transformed random variable $Y=F_X(X)$ follows a standard uniform distribution (i.e., $Y\sim U(0, 1)$). In practice, a PIT value is often computed for each observation $y_i$ as a proportion of its corresponding predictive draws $y_i^{\text{rep}}$ that are less than or equal to the observed value. Formally:

\begin{equation}
    \text{PIT}_i=\frac{1}{S}\sum_{s=1}^SI(y_i^{\text{rep}, s}\le y_i),
\end{equation}
where $S$ is the number of predictive draws and $I(\cdot)$ is an indicator function. After the PIT values are computed, their empirical cumulative distribution function (ECDF) is plotted. If the model is well calibrated, the ECDF of the PIT values should be close to a straight line from 0 to 1, as this indicates that the PIT values follow a standard uniform distribution.

Figure \ref{fig:ppc_km_overlay_observed} displays the Kaplan-Meier overlay plot. The plot shows the Kaplan-Meier estimator \citep{kaplan1958} of the survival function $S(t)$ for the observations. The estimator is defined as

\begin{equation}
    \widehat{S}(t)=\prod_{i:t_i\le t}\left( 1 - \frac{d_i}{n_i} \right),
    \label{eq:kaplan_meier}
\end{equation}
where $t_i$ are distinct event times, $d_i$ is the number of individuals who experience the event at time $t_i$, and $n_i$ is the number of individuals at risk just before time $t_i$. For predictive draws, the plot simply shows the empirical complementary cumulative distribution functions (CCDF), which the Kaplan-Meier curve of the observations can be compared against.

\begin{figure}[t]
    \centering % Center the entire figure content
    \begin{subfigure}{0.49\textwidth}
        \centering
        \includegraphics[width=\linewidth]{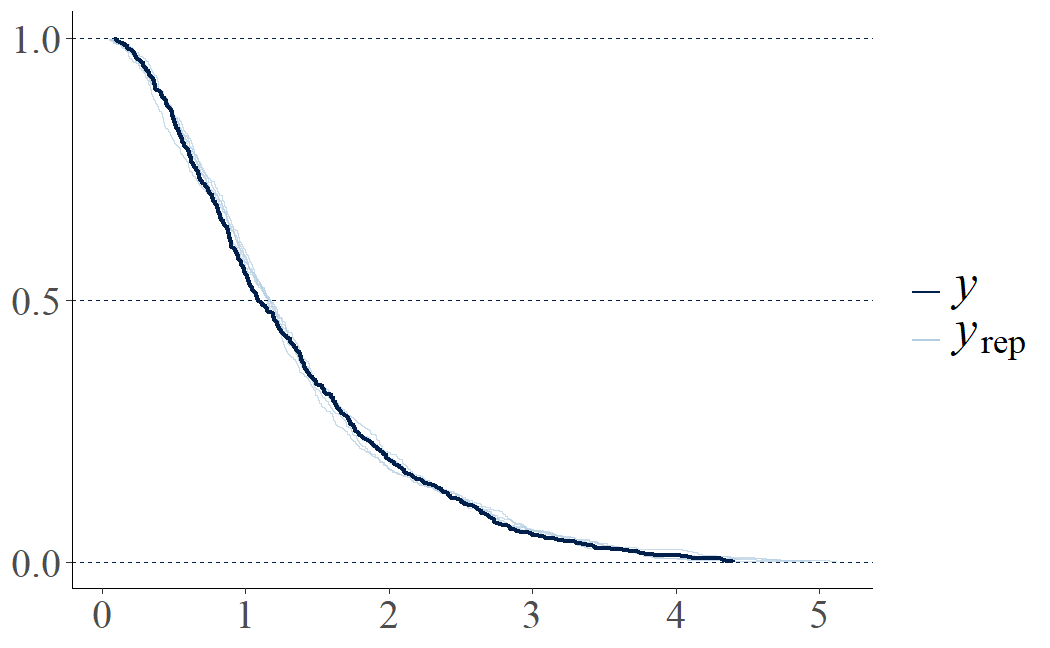}
        \caption{In the above figure, left-truncation is not taken into account. That is, all individuals that have not experienced the event before time point $t$ are considered to be at risk at time point $t$.}
        \label{fig:a_ppc_km_overlay_left_truncation_problem}
    \end{subfigure}
    \hfill
    \begin{subfigure}{0.49\textwidth}
        \centering
        \includegraphics[width=\linewidth]{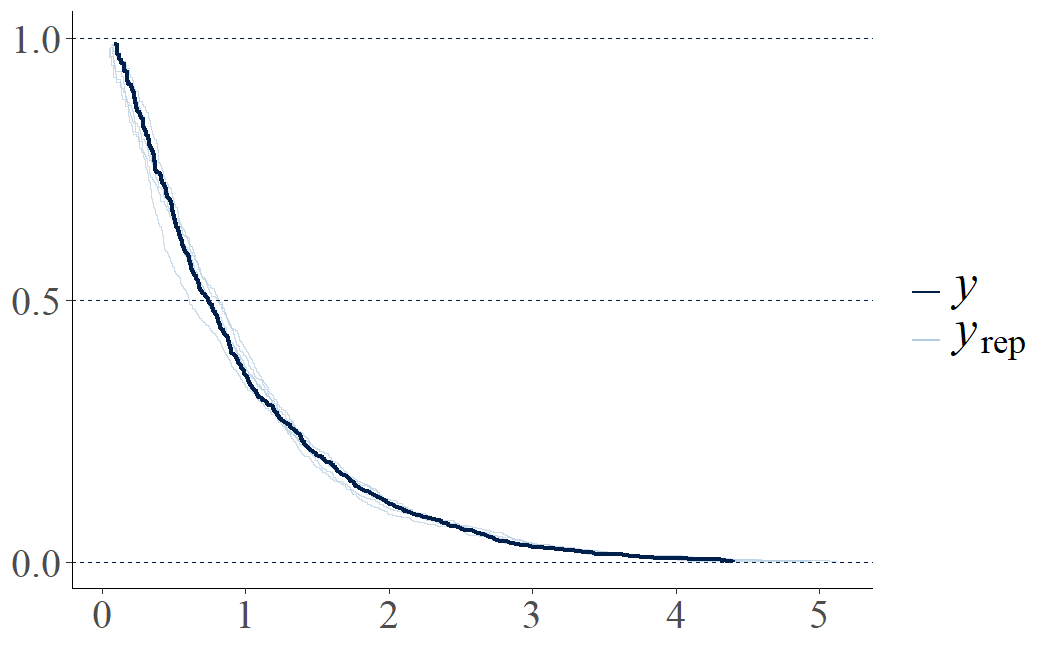}
        \caption{In the above figure, left-truncation is taken into account by considering individuals to be at risk only after they have entered the study.\\~}
        \label{fig:b_ppc_km_overlay_left_truncation_solution}
    \end{subfigure}
    % The main figure caption
    \caption{Two different Kaplan-Meier overlay plots for the same data set where some data points are left-truncated. a) Left-truncation is not taken into account. b) Left-truncation is taken into account.}
    \label{fig:ppc_left_truncation}
\end{figure}

There is a small detail in Kaplan-Meier overlay plots that has received less attention in the software implementations. When the data set contains delayed entries, that should be taken into account in the Kaplan-Meier overlay plot. Individuals who enter the study late are conditioned on the fact that they have not experienced the event before their entry time point, which means that they do not represent the population at the beginning of the study. Therefore, these individuals cannot be considered at risk before entering the study. That is, $n_i$, which represents the number of individuals at risk at the time point $t_i$ in Equation \ref{eq:kaplan_meier}, should not count individuals who have entered the study after the time point $t_i$. Otherwise, $n_i$ would be too high, which, as seen in the equation, would result in overestimated survival probabilities at early time points. The impact of this mistake is illustrated in Figure \ref{fig:ppc_left_truncation}.

\subsection{Predictive checking of parametric survival models in the presence of censored event times}

As stated previously, the predictive model checking requires the existence of a generative model. In case the data set contains censored event times, creating the full generative model requires a model for the censoring process. However, this is tricky to create in practice, as there may be multiple different reasons for censoring. In this subsection, we do not model the censoring process, but we assume it to be independent from the survival process.

\begin{figure}[t]
    \centering
    \begin{subfigure}{0.49\textwidth}
        \centering
        \includegraphics[width=\linewidth]{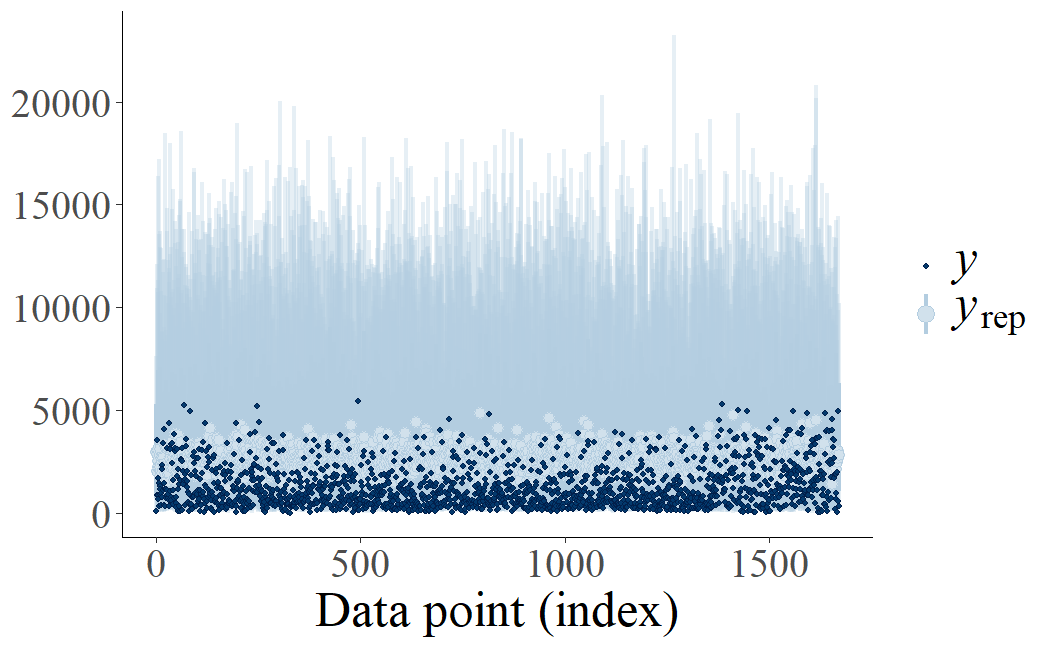}
        \caption{The intervals plot in the figure above shows that the
medians of the predictive draws are generally above the observations, and the central intervals of the predictive draws continue much further than there are observations.}
        \label{fig:a_ppc_intervals_censored}
    \end{subfigure}
    \hfill
    \begin{subfigure}{0.49\textwidth}
        \centering
        \includegraphics[width=\linewidth]{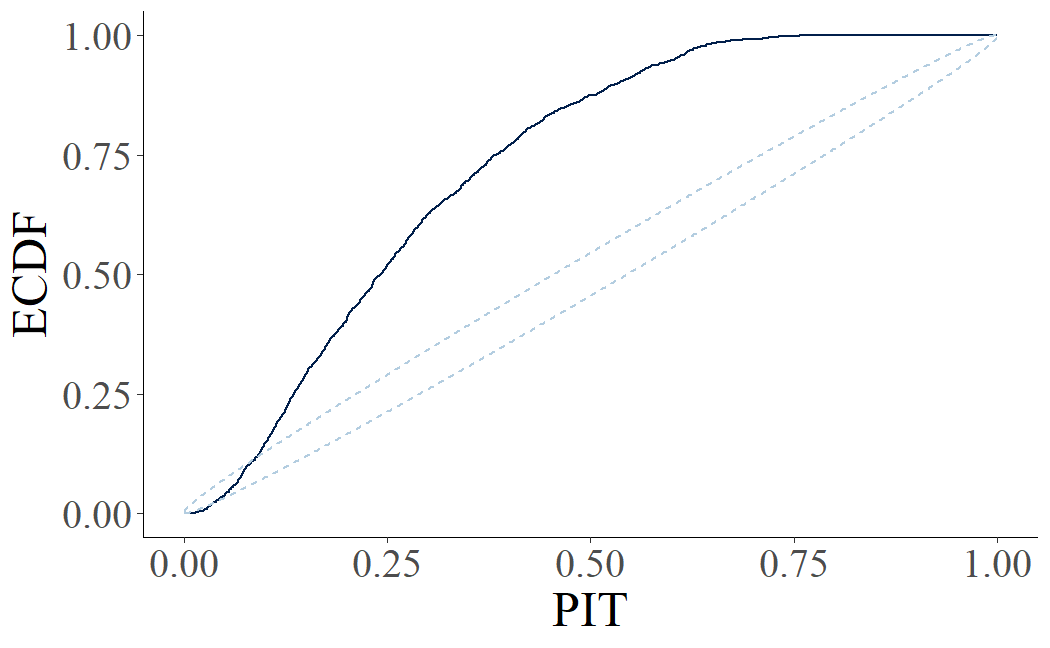}
        \caption{The PIT-ECDF plot in the figure above shows that the PIT-ECDF curve exceeds the confidence bands by a wide margin.\\~}
        \label{fig:b_pit_ecdf_censored}
    \end{subfigure}
    \caption{Two diagnostic plots for the same data that contains a lot of censored event times. a) Intervals plot. b) PIT-ECDF plot.}
    \label{fig:ppc_diagnostic_censored}
\end{figure}

When the censoring process is not modelled, some diagnostic plots become useless. Examples of these types of plot are the intervals plot and the PIT-ECDF plot, which are shown in Figure \ref{fig:ppc_diagnostic_censored}. These two diagnostic plots always indicate a lack of fit when there are censored event times in the data and the censoring process is not modelled. Therefore, these diagnostic plots do not provide much value as a predictive model check in this kind of case, and we recommend not using them unless the censored event times are dealt with.

One method of dealing with censored event times is imputation. Imputation means that after fitting a model, we draw missing event times from the posterior predictive distribution. If the censored event times are imputed, the intervals plot and the PIT-ECDF plot become useful. However, it is advisable to indicate imputed data points in some way in the diagnostic plots. For example, they can be plotted with a different colour than the observed data points.

\begin{figure}[p]
  \begin{center}
    \includegraphics[width=0.49\textwidth]{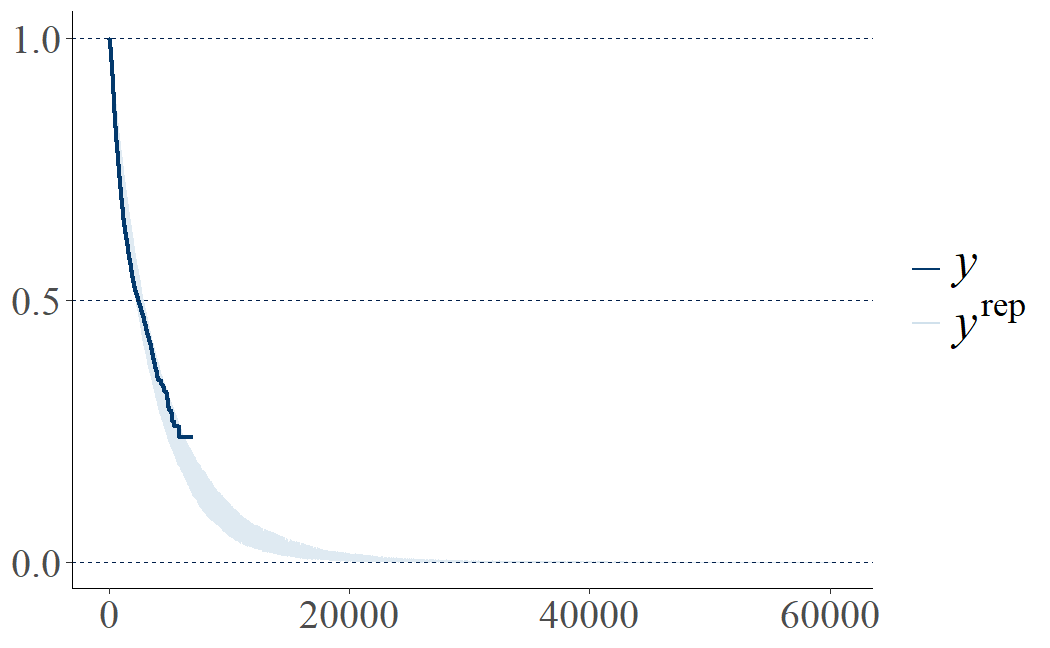}
    \caption{Kaplan-Meier overlay plot when all predictive draws are shown.}
    \label{fig:ppc_km_overlay_extrapolation_problem}
  \end{center}
\end{figure}

\begin{figure}[p]
  \begin{center}
    \includegraphics[width=0.49\textwidth]{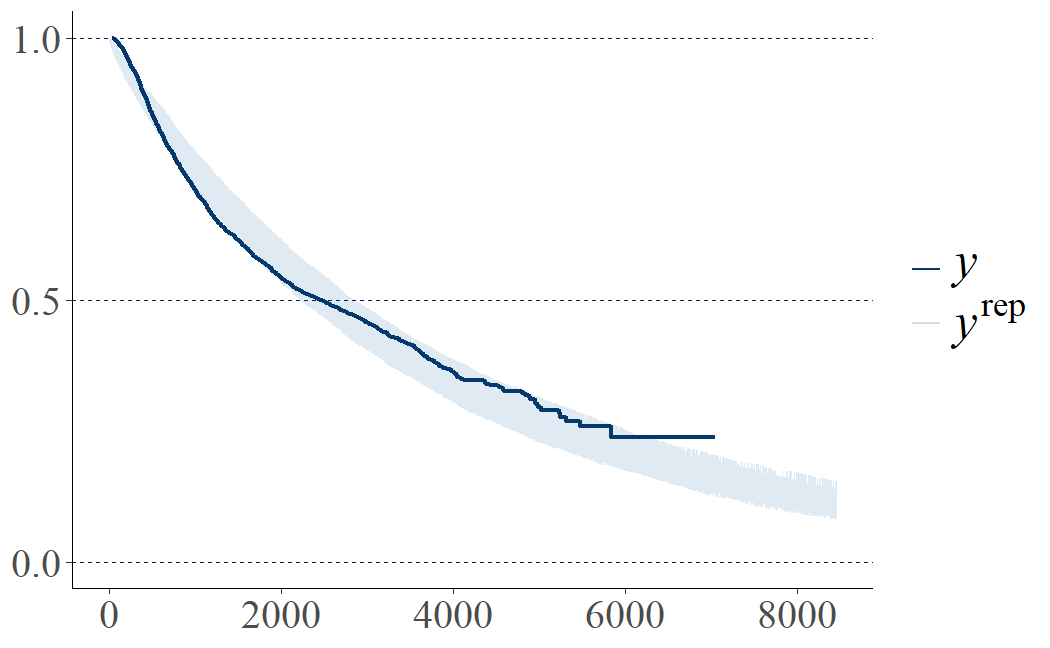}
    \caption{Kaplan-Meier overlay plot when predictive draws are shown at most 20\% beyond the furthest observation.}
    \label{fig:ppc_km_overlay_extrapolation_solution}
  \end{center}
\end{figure}

\begin{figure}[p]
  \begin{center}
    \includegraphics[width=0.49\textwidth]{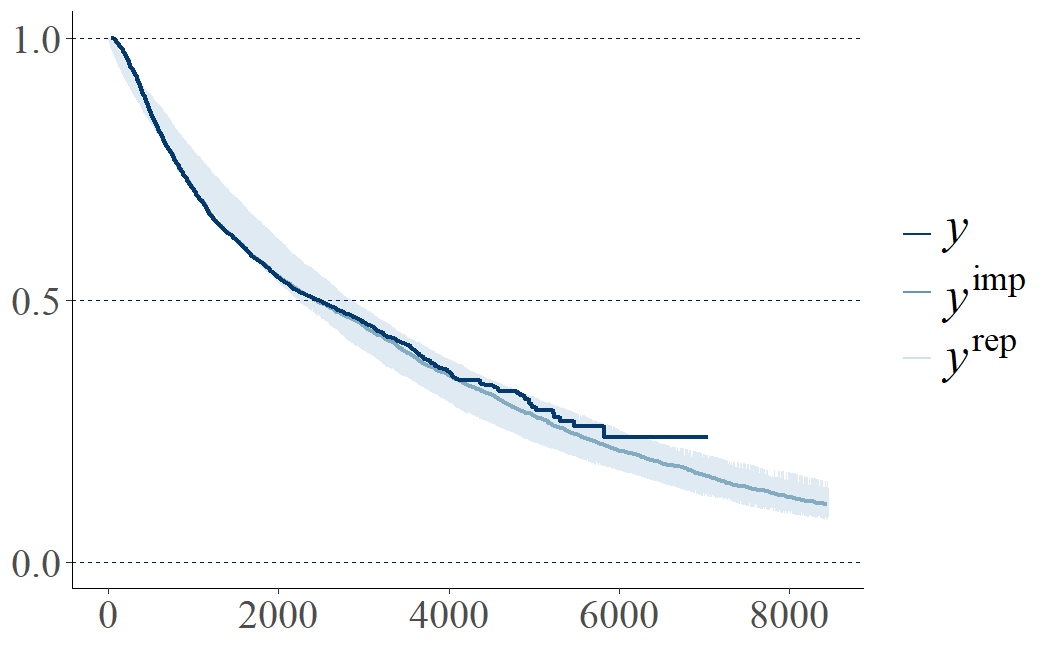}
    \caption{The same as Figure \ref{fig:ppc_km_overlay_extrapolation_solution} except that a Kaplan-Meier survival curve, which includes imputed censored event times, is plotted alongside the survival curve for the observations.}
    \label{fig:ppc_km_overlay_imputation}
  \end{center}
\end{figure}

When the data set contains censored event times and the censoring process is not modelled, Kaplan-Meier overlay plots remain useful for predictive model assessment. However, it is often useless to show the full empirical CCDF estimates of the predictive draws if a large portion of event times are censored. This is demonstrated in Figures \ref{fig:ppc_km_overlay_extrapolation_problem}, \ref{fig:ppc_km_overlay_extrapolation_solution}, and \ref{fig:ppc_km_overlay_imputation}. The figures show three different Kaplan-Meier overlay plots for the same data set that contains a lot of censored event times.

Figure \ref{fig:ppc_km_overlay_extrapolation_problem} visualises what can happen if the full empirical CCDF curves of the predictive draws are shown. In the figure, the predictive draws continue 10 times further than the observations, and the largest predictive draws are at around 60 000 days, which converts to roughly 165 years. This is more than the longest human lifetime ever recorded. It is unreasonable to expect to be able to extrapolate so far. In addition, plotting the long tail for the predictive draws does not provide useful information as a predictive model check. For example, if we have a study that ends at some point in time, the observations obviously stop when the study ends. Thus, we cannot judge a model to be a poor fit if it has a longer tail than the observations.

Figure \ref{fig:ppc_km_overlay_extrapolation_solution} shows an improved version of the Kaplan-Meier overlay plot in which the extrapolation is shown 20\% beyond the furthest observed time point. In this visualisation, the main focus is on the observed data, which is the most interesting part of the figure. The zoomed plot reveals that the Kaplan-Meier survival curve for the observations deviates from the empirical CCDF estimates for the predictive draws, which would have been difficult to notice in Figure \ref{fig:ppc_km_overlay_extrapolation_problem}.

Figure \ref{fig:ppc_km_overlay_imputation} further improves the Kaplan-Meier overlay plot by considering censored event times as missing data, imputing them, and plotting the Kaplan-Meier survival curve that includes imputed data points in addition to the survival curve of the observations. This clearly illustrates when the censored event times start to affect the model.

\subsection{Predictive checking of the Bernoulli model}

In contrast to other survival models, the Bernoulli model has a binary outcome. For binary outcomes, a common mistake when performing a predictive model check is the use of bar graphs. Bar graphs are plots that show bars corresponding to frequencies of observed values and point intervals of predictive means and centre quantiles. According to \cite{sailynoja2025}, bar graphs are often meaningless as a predictive model check for binary data, since even the simplest intercept only models can perfectly model the proportions of the values, and bar graphs do not allow inspection of any other aspects, such as under- and overconfidence.
\begin{figure}[t]
  \begin{center}
    \includegraphics[width=0.49\textwidth]{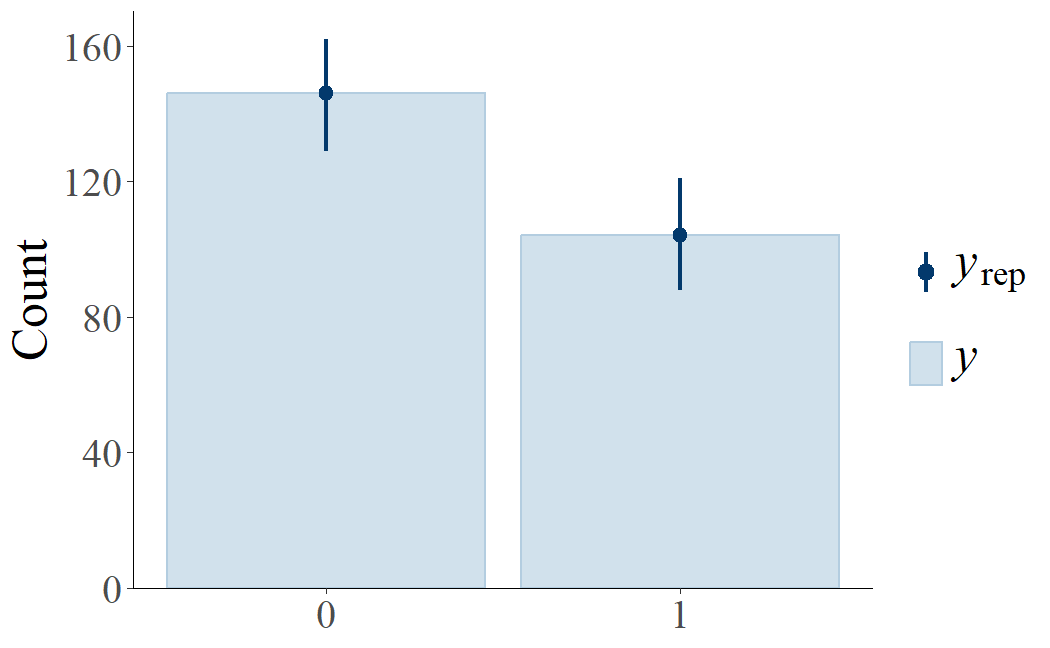}
    \caption{Example of a bar graph for binary target.}
    \label{fig:ppc_bars}
  \end{center}
\end{figure}

To create a more comprehensive model check for binary outcomes, \cite{sailynoja2025} recommend the PAV-adjusted calibration plots applying the work by \cite{dimitriadis2021}. The idea of PAV-adjusted calibration plots is to compute conditional event probabilities (CEPs) using the posterior predictive means for each observation and then to form consistency bands that show how much variation should be expected from a perfectly calibrated model. CEPs denote the probability that the event occurs given that the model has predicted some probability. In a perfectly calibrated model, the CEPs are equal to the predicted probabilities. The CEPs are computed in the PAV-adjusted calibration using an algorithm called pool adjacent violators (PAV) \citep{ayer1955}, which the calibration plot is named after.
\begin{figure}[p]
  \begin{center}
    \includegraphics[width=0.65\textwidth]{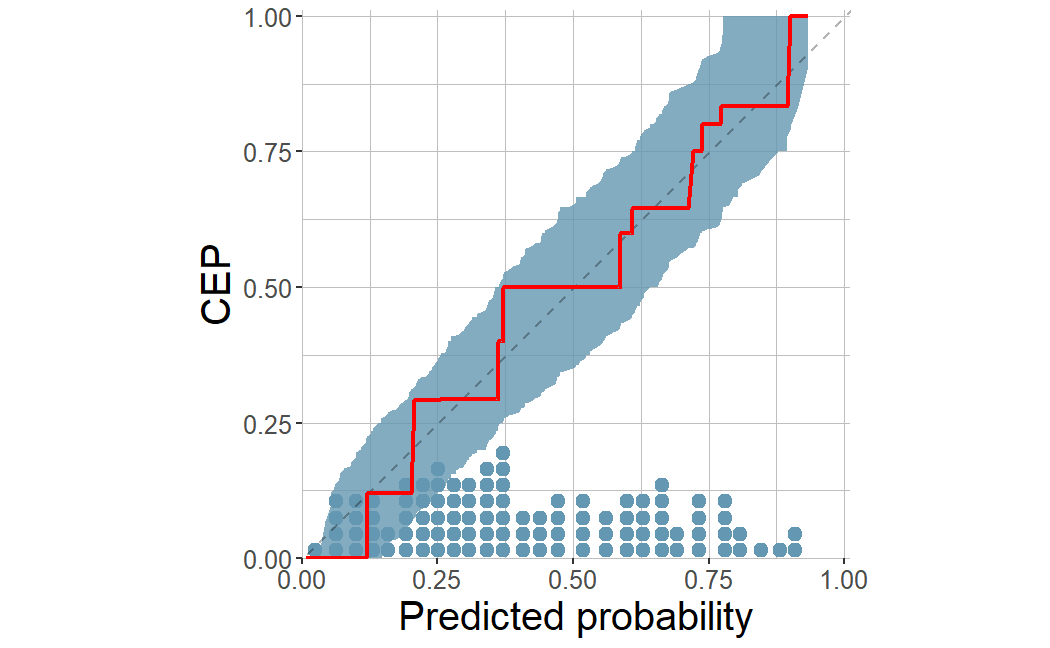}
    \caption{Example of a PAV-adjusted calibration plot. In the figure, red line indicates the calibration curve for the model, dashed diagonal line indicates the line of perfect calibration, blue ribbon indicates the consistency band for the calibration curve, and blue dots indicate where most predicted probabilities lie. A calibration curve that outpasses the consistency band is an indication of miscalibration.}
    \label{fig:ppc_calibration_pava}
  \end{center}
\end{figure}

\begin{figure}[p]
  \begin{center}
    \includegraphics[width=0.65\textwidth]{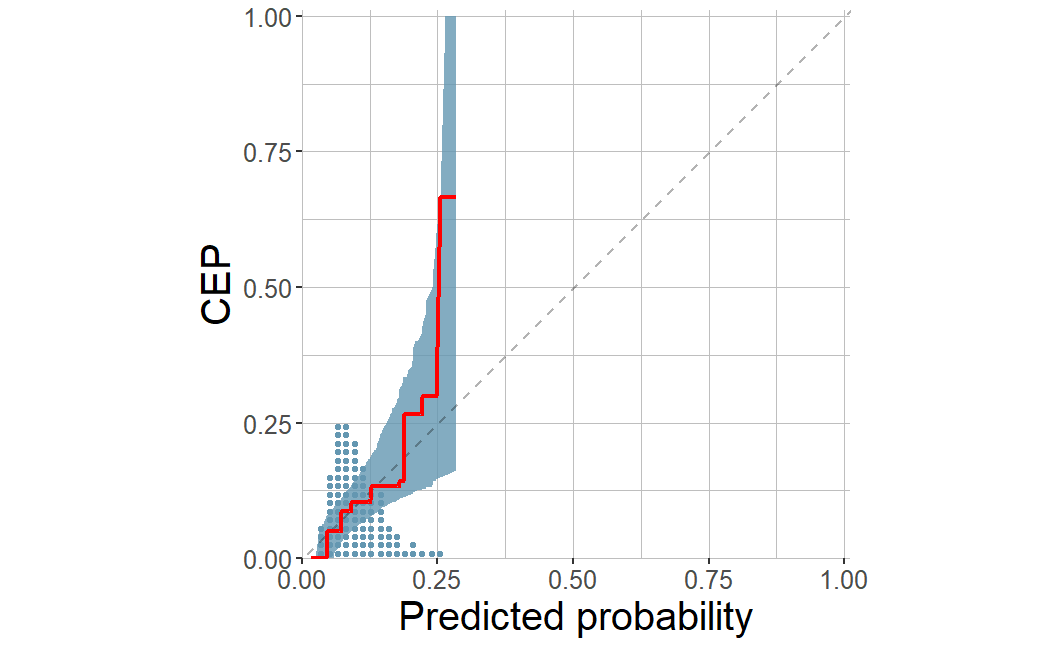}
    \caption{PAV-adjusted calibration plot in the case where most probabilities are near zero.}
    \label{fig:ppc_calibration_pava_unbalanced}
  \end{center}
\end{figure}

In case the binary data contain one value a lot more than the other, a fitted model predicts a lot of low probabilities and very few high probabilities, or the other way around. A situation like this is presented in Figure \ref{fig:ppc_calibration_pava_unbalanced}. The small blue dots in the figure indicate that most predictions are near zero.

\begin{figure}[p]
  \begin{center}
    \includegraphics[width=0.65\textwidth]{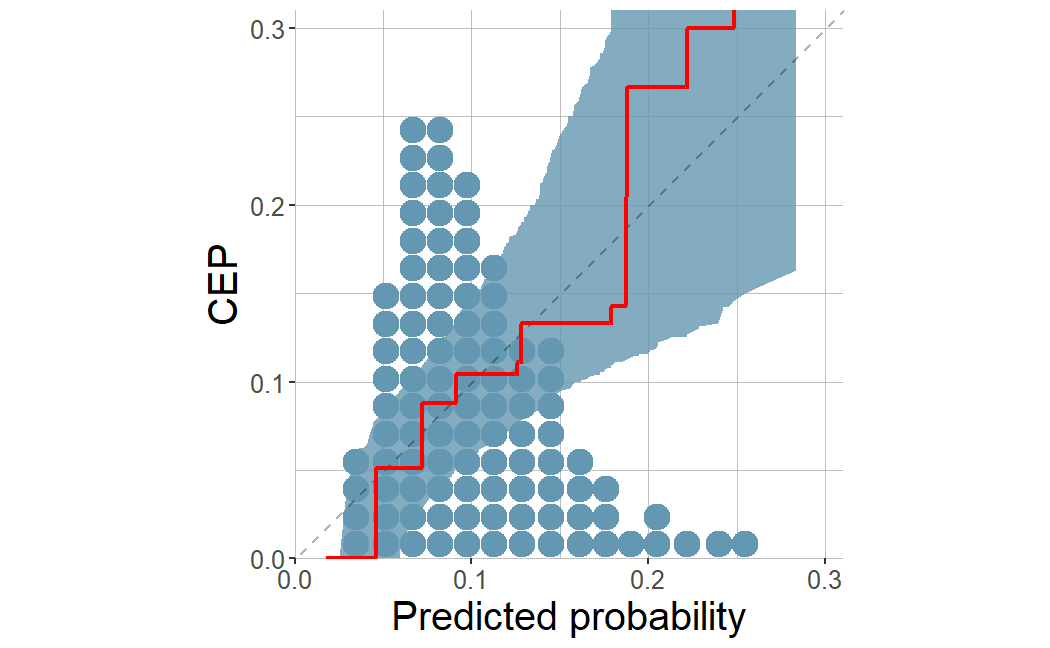}
    \caption{Zoomed PAV-adjusted calibration plot in the case where most probabilities are near zero.}
    \label{fig:ppc_calibration_pava_unbalanced_zoomed}
  \end{center}
\end{figure}

In this kind of case, we also recommend inspecting a zoomed image that focuses on the area where most of the small blue dots lie. It is possible that the zoomed plot reveals things that would have been difficult to spot in the larger image. A zoomed version of the previous image is shown in Figure \ref{fig:ppc_calibration_pava_unbalanced_zoomed}.

\subsection{Predictive model checks suitable for any survival model}

It is also possible to create predictive model checks that are suitable for any survival model in the presence and absence of censored event times regardless of the target variable of the model. These types of predictive model checks require compromises that can potentially lose some information.

A simple way to perform a universal predictive model check for a survival model is to dichotomise the prediction so that we use the model to predict whether or not an event happens before some fixed time point (e.g., 5 years). This makes the outcome binary regardless of the original survival model. Thus, after dichotomisation, the previously presented recommendations for performing a predictive model check for the Bernoulli model can be applied.

The aforementioned approach can be extended to perform a more comprehensive predictive model check. The more comprehensive approach that we suggest is to discretise time to some fixed length interval and for each interval perform a prediction on whether the event happens in that interval or not. In this case, the outcome is binary for each interval, and thus the predictive model check can be performed by plotting multiple PAV-adjusted calibration plots, one corresponding for each interval.

\section{Predictive comparison of survival models}
\label{sec:predictive_model_comparison}

To compare predictive performance between models, a common approach is to compute predictive densities using log scores \citep{geisser1979, bernardo1994, gneiting2007} as a utility. Furthermore, \cite{vehtari2017} define expected log-predictive densities as a measure of predictive performance for a new data set of $n$ observations taken one at a time

\begin{equation}
\mathrm{elpd}(M_k \mid y) = \sum_{i=1}^n \int p_t(\tilde{y}_i) \log p_k(\tilde{y_i} \mid y) d\tilde{y}_i,
\end{equation}
where $p_t(\tilde{y}_i)$ refers to the distribution of the true data generation process for $\tilde{y}_i$, and $p_k(\tilde{y}_i \mid y) = \int p_k(\tilde{y}_i \mid \theta) p_k(\theta \mid y) d\theta$ denotes the posterior predictive distribution of the model $M_k$. The $p_t(\tilde{y}_i)$'s are unknown, which means that elpd can only be approximated. One way to perform the approximation is leave one out cross-validation (LOO-CV):

\begin{equation}
\mathrm{elpd_{LOO}}(M_k \mid y)=\sum_{i=1}^n \log p_k(y_i \mid y_{-i})=\sum_{i=1}^n \log\int p_k(y_i \mid \theta) p_k(\theta \mid y_{-i}) d\theta.
\label{eq:elpd_loo}
\end{equation}

The state-of-the-art method for computing estimates for $\mathrm{elpd_{LOO}}$ \citep{vehtari2017} uses Pareto smoothed importance sampling (PSIS) \citep{Vehtari-Simpson-Gelman-etal:2024}. In practice, the PSIS-LOO CV based model comparison can be performed using the functions in the loo R package \citep{vehtari2024_loo}. The \texttt{loo()} function can be used to compute the elpd estimate $\widehat{\mathrm{elpd}}$ for each model. The \texttt{loo\_compare()} function can then compare the elpd estimates between models. We denote the differences in the elpd estimates between models by $\Delta\widehat{\mathrm{elpd}}$. For the best model, $\Delta \widehat{\mathrm{elpd}} = 0$, and for the other models, $\Delta\widehat{\mathrm{elpd}}$ closer to zero is better. \texttt{loo\_compare()} also reports the standard error related to the comparison. According to  \cite{vehtari2017}, the standard error of $\mathrm{\widehat{elpd}}$ is defined as

\begin{equation}
\mathrm{se(\widehat{elpd}_{LOO})}=\sqrt{nV_{i=1}^n(\mathrm{\widehat{elpd}_{LOO,i}})},
\end{equation}
where $n$ is the number of data points. As justified by \cite{Sivula-Magnusson-Vehtari:2025}, the standard error can be used to quantify the uncertainty in the comparison of models. The standard error related to the comparison of models A and B is

\begin{equation}
    \mathrm{se(\Delta\widehat{elpd}_{LOO})}=\sqrt{nV_{i=1}^n(\mathrm{\widehat{elpd}_{LOO,i}^A}-\mathrm{\widehat{elpd}_{LOO,i}^B})}.
\label{eq:se_comp}
\end{equation}
If $\Delta\widehat{\mathrm{elpd}}$ is less than two standard errors from zero, the model is considered indistinguishable in predictive performance compared to the best model \citep{riha2024}.

Although elpd estimates computed with PSIS-LOO CV are often excellent for model comparison, there are some important limitations to them and things to keep in mind. This section delves into the applicability and limitations of elpds in survival models. The section especially focuses on how censored event times that are not included in the generative model affect the model comparison.

Since the PH and AFT models have the same target variable when fitted with brms, they behave similarly in the predictive model comparison. However, the Bernoulli model has a different target variable than the PH and AFT models and is a very different model in many other ways as well, so the recommendations are different when the Bernoulli model is included in the comparison. Even if the Bernoulli model is not included, having censored event times in the data affects the model comparison. Therefore, the cases that we discuss in this section are 1. PH and AFT models in the absence of censored event times, 2. PH and AFT models in the presence of censored event times, and 3. PH, AFT, and Bernoulli models.

\subsection{Predictive comparison of PH and AFT models in the absence of censored event times}

\begin{figure}[t]
    \centering
    \begin{subfigure}{0.49\textwidth}
        \centering
        \includegraphics[width=\linewidth]{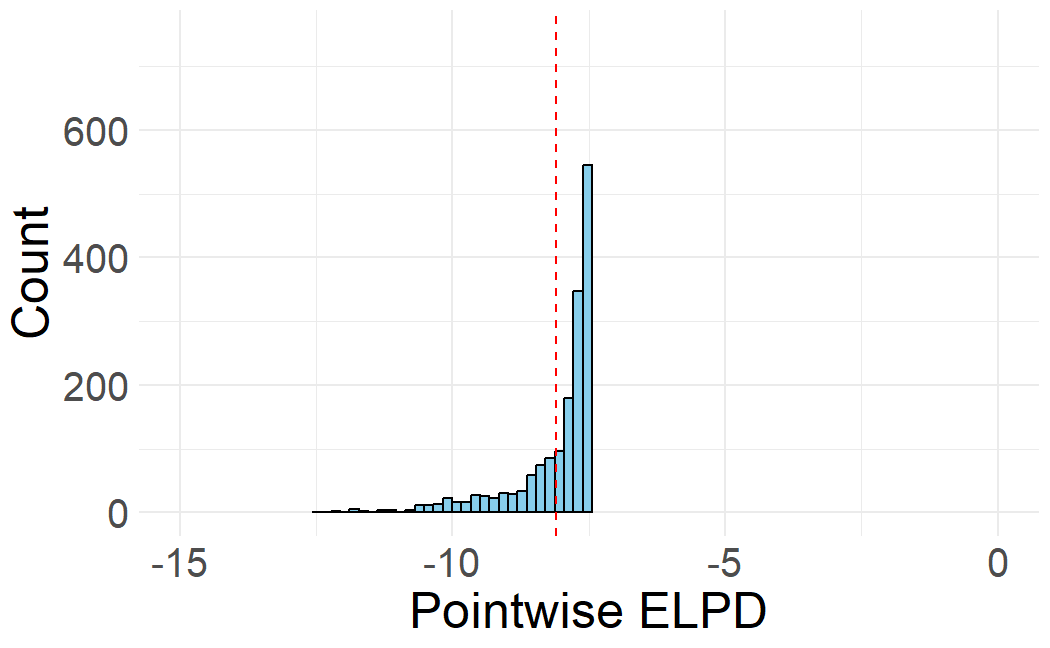}
        \caption{Histogram of pointwise elpds when the event time is in days.}
        \label{fig:a_elpd_hist_days_observed}
    \end{subfigure}
    \hfill
    \begin{subfigure}{0.49\textwidth}
        \centering
        \includegraphics[width=\linewidth]{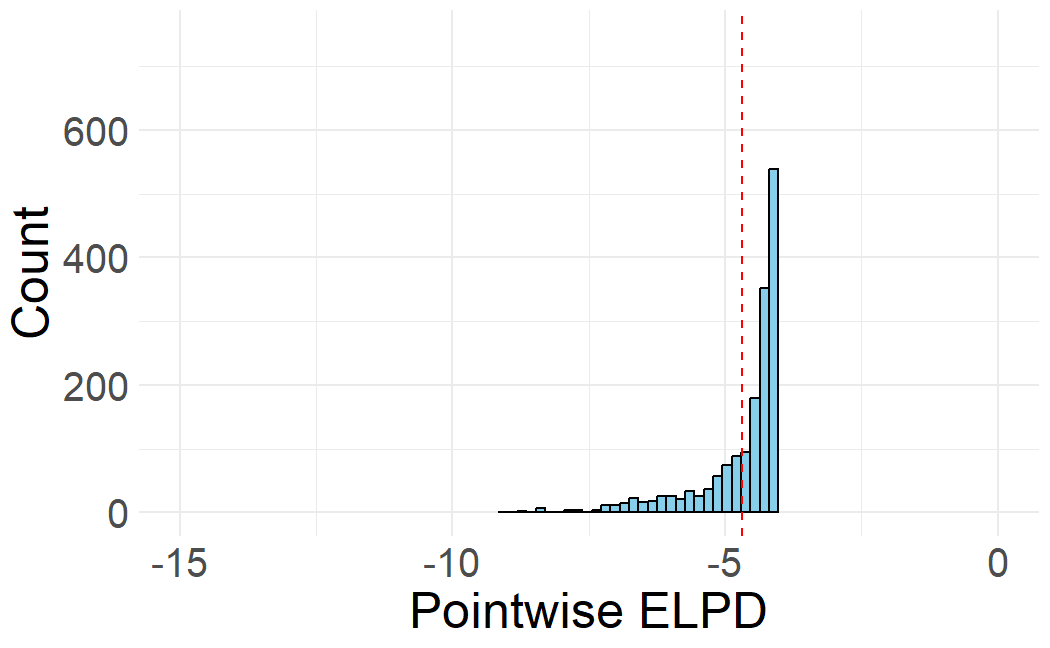}
        \caption{Histogram of pointwise elpds when the event time is in months.}
        \label{fig:b_elpd_hist_months_observed}
    \end{subfigure}
    \caption{Two elpd histograms obtained from two identical Weibull AFT models fitted to nearly identical data sets that do not contain censored event times. The only difference between the data sets is the time scale of the target variable. a) The time scale of the target variable is days. b) The time scale of the target variable is months.}
    \label{fig:elpd_hist_observed}
\end{figure}

The simplest case of survival model comparison is to compare AFT models and PH models when the data set does not contain censored event times. In this kind of case, applying the PSIS-LOO CV with the usual recommendations is sufficient, as long as the time scales of the target variables are the same in all models. Figure \ref{fig:elpd_hist_observed} illustrates what happens if the time scales of the target variables differ between the models. The figure shows histogram bars for pointwise elpd counts and red dashed lines indicating the elpd means for two cases that only differ in the time scale of the target variable. As can be seen in the figure, a change in the target variable's time scale shifts the pointwise elpds making a comparison across time scales invalid.

\subsection{Predictive comparison of PH and AFT models in the presence of censored event times}
\label{subsec:predictive_comparison_ph_aft_censoring}

When the data set contains censored event times, the predictive model comparison becomes more complicated. In case the data point $y_i$ is censored, $p_k(y_i \mid \theta)$ in Equation \ref{eq:elpd_loo} is interpreted in terms of the true event time $y_i^\ast$ as
\begin{enumerate}
\item $p_k(y_i^\ast > a \mid \theta)$, if $y_i$ is right-censored at $a$;
\item $p_k(y_i^\ast < b \mid \theta)$, if $y_i$ is left-censored at $b$; and
\item $p_k(a < y_i^\ast < b \mid \theta)$, if $y_i$ is interval-censored with endpoints $a$ and $b$.
\end{enumerate}
This means that pointwise elpds for censored data points are not densities, but probabilities. The phenomenon can be clearly illustrated when computing pointwise elpds for two models, which are otherwise the same, but the time scales of the event times are different. Now, pointwise elpds stay the same for the censored data points but change for the observed ones.

\begin{figure}[t]
    \centering % Center the subfigures
    \begin{subfigure}{0.49\textwidth}
        \centering
        \includegraphics[width=\linewidth]{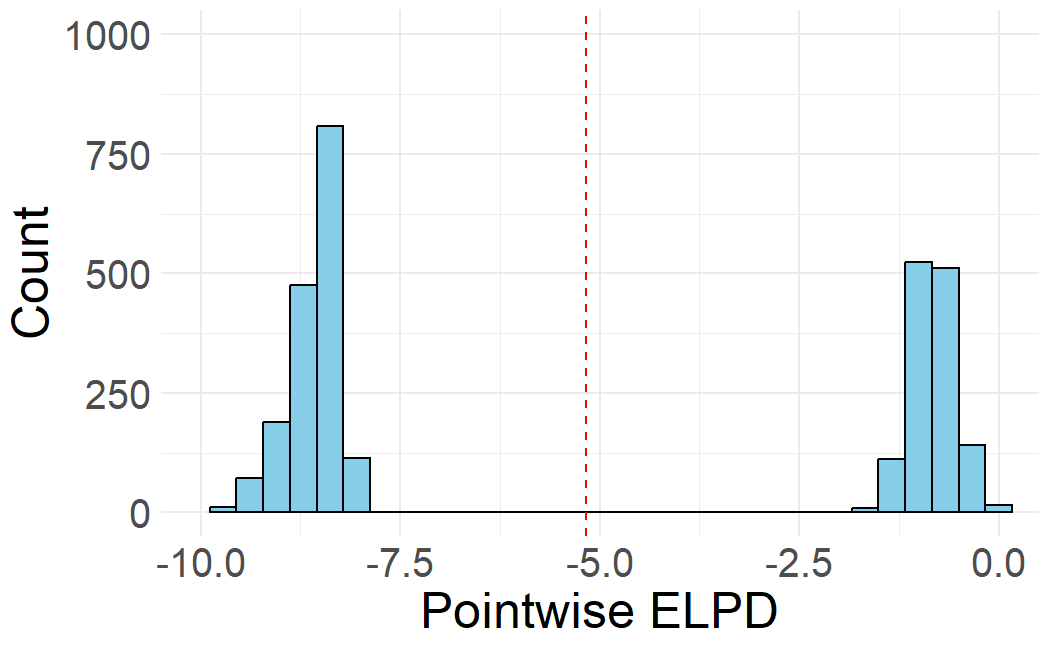}
        \caption{Histogram of pointwise elpds when the event time is in days.}
        \label{fig:a_elpd_hist_days}
    \end{subfigure}
\hfill
    \begin{subfigure}{0.49\textwidth}
        \centering
        \includegraphics[width=\linewidth]{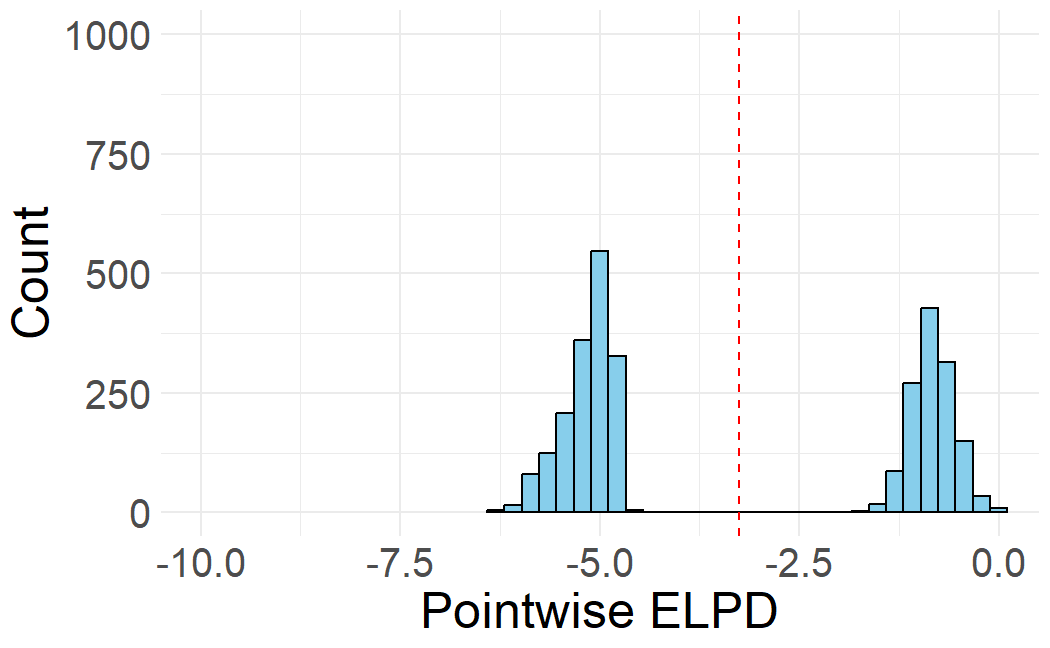}
        \caption{Histogram of pointwise elpds when the event time is in months.}
        \label{fig:b_elpd_hist_months}
    \end{subfigure}
    \caption{Two elpd histograms obtained from two identical Weibull AFT models fitted to nearly identical data sets where a significant portion of event times are censored. The only difference between the data sets is the time scale of the target variable. a) The time scale of the target variable is days. b) The time scale of the target variable is months.}
    \label{fig:elpd_hist_censored}
\end{figure}

To create an example, we fitted a Weibull AFT model on a data set where the event time is in days, and a significant portion of the event times are right-censored. After that, we computed pointwise elpds and plotted their histogram in Figure \ref{fig:a_elpd_hist_days}. There are two clear clusters in the figure. The right cluster contains the pointwise elpds for the censored data points, and the left cluster has the pointwise elpds for the non-censored data points.

We also fitted the Weibull AFT model to a data set that is otherwise the same as in the previous case, but the time scales of the event times are changed to months. The histogram of pointwise elpds is shown in Figure \ref{fig:b_elpd_hist_months}. The figure shows that the location of the censored elpd cluster (right cluster) remains the same but the location of the non-censored elpd cluster (left cluster) is now closer to the censored elpd cluster.

The change in the distance between the clusters as a consequence of the time scale change is problematic because an elpd is a depiction of how easy a data point is to predict. The fact that elpd stays the same for some data points and changes for others indicates a change in the relative difficulty of predicting the data points. The issue can be eliminated by avoiding mixing probabilities and densities.

It is not possible to convert probabilities into densities, but it is possible to convert densities into probabilities. Once the densities have been converted into probabilities, the PSIS-LOO CV can be applied without issues. The conversion can be performed by integrating the densities over different time intervals. In general, the shorter the time intervals are, the less information is lost. However, when choosing the interval length in practice, one must evaluate the tradeoff between preserving information and keeping the computational cost manageable, since shorter intervals also mean higher computational cost. A good starting point for deciding the interval length is to consider the data measurement resolution, as intervals shorter than the smallest possible time increment in the data do not add additional value compared to intervals that match the smallest possible time increment. For example, if a new data point is created every month, it is not useful to discretise the time into days. If using intervals that match the smallest possible time increment comes with overwhelming computational cost, one can gradually increase the interval length bearing in mind that each increase potentially loses some information. In order to consider the interval length to be information preserving, one should be able to assume that the probability density function remains approximately constant within each interval.

\subsection{Predictive comparison of PH, AFT, and Bernoulli models}

The Bernoulli model is very different from the PH and AFT models in multiple ways. The Bernoulli model has a binary target variable and applies a long data-format, which contains multiple data points for each subject. In order to compare the predictive performance of the Bernoulli model against PH and AFT models, we must take these differences into account.

In Section \ref{subsec:predictive_comparison_ph_aft_censoring}, we recommended converting densities to probabilities, when comparing AFT and PH models in the presence of censored event times. However, the Bernoulli model utilises discretised time, and thus the model functions in probabilities as a default. Therefore, the time in PH and AFT models should be discretised to the same intervals as in the Bernoulli model regardless of whether there are censored event times in the data set. Otherwise, we would be comparing densities with probabilities.

Adjusting for multiple data points for each patient in the cross-validation of the Bernoulli model is straightforward. We must simply leave out all the data points corresponding to one subject and perform the prediction for the left out data points. 

The computational comparison certainly requires more effort when the Bernoulli model is included in the comparison. However, a full computational comparison is not always necessary. This is because unlike the PH and AFT models, the Bernoulli model allows for the inclusion of time-dependent variables. Therefore, if the data show clear time-dependent effects for some variables, the Bernoulli model can often be declared to be the superior model without any computational comparison.

\section{Case study: Gastrointestinal stromal tumour}
\label{sec:case_study}

In this section, we present a case study that demonstrates the methodology covered in the previous sections of this paper. The case study is based on a real data analysis task that focuses on the risk of recurrence of a gastrointestinal stromal tumour after surgery. The analysis used a data set that was collected over multiple decades in a collaborative effort of multiple research centres around the world.

However, we are not allowed to publish results from the real data set in this paper. Thus, we created a simulated data set that attempts to mimic the original data set and data sets that have been used in previous GIST studies \citep{joensuu2012, joensuu2014}. As the data are synthetic, the purpose of this section is not to focus on the results. Rather, the section aims to focus on the model checking methodology that is presented in earlier sections of this paper and is now applied in practice. 

\begin{figure}[t]
    \centering
    \begin{subfigure}{0.65\textwidth}
        \centering
        \includegraphics[width=\linewidth]{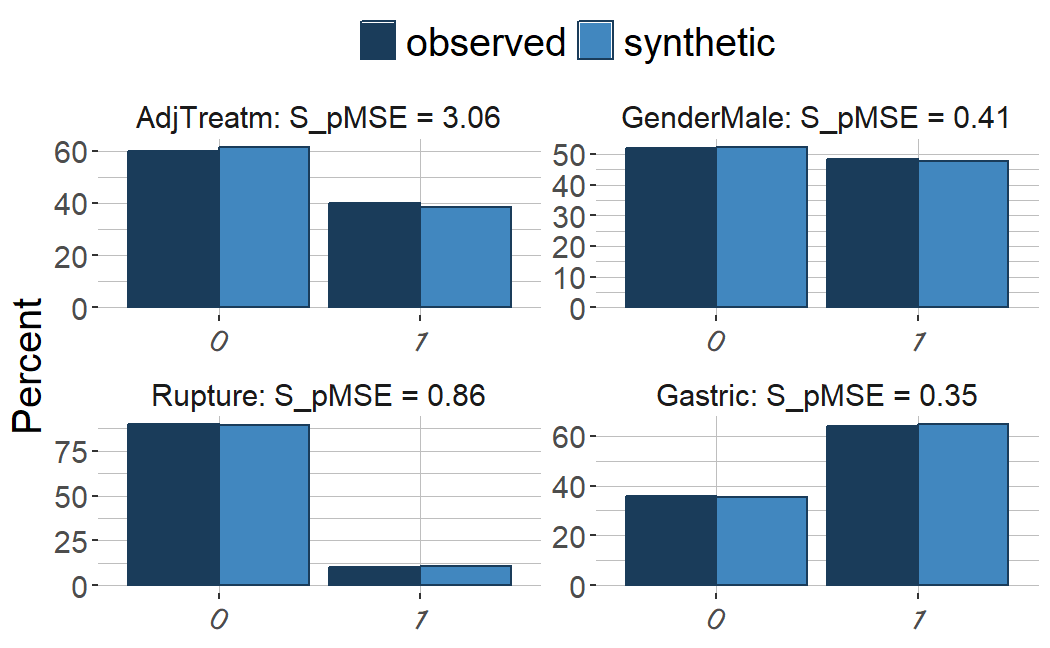}
    \end{subfigure}
    \begin{subfigure}{0.65\textwidth}
        \centering
        \includegraphics[width=\linewidth]{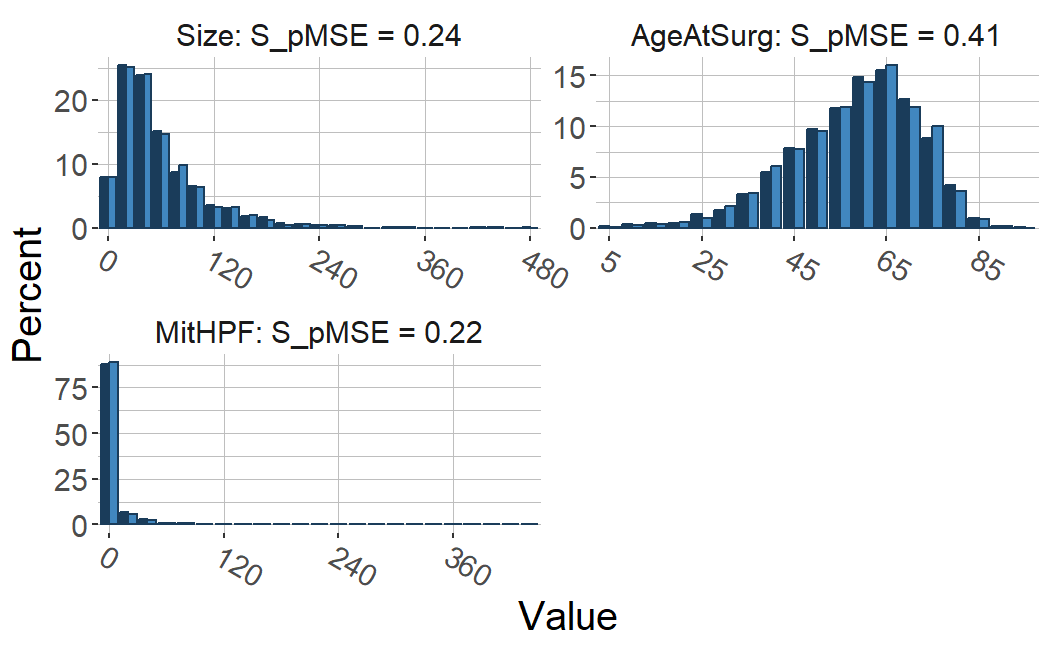}
    \end{subfigure}
    \caption{Distributions of variables in synthetic data and original data.}
    \label{fig:syndata}
\end{figure}

The synthetic data set was generated using the following procedure. First, we removed data points with missing values from the original data set and fit the Bernoulli model to it. Then, we simulated covariates. We chose only covariates that were used by \cite{joensuu2012}, although the original data set had other covariates as well. The selected covariates were Size, AgeAtSurg, MitHPF, GenderMale, Rupture, and Gastric. In addition, we included information on whether the patient received adjuvant treatment or not in a variable AdjTreatm. We created the synthetic data set consisting of these variables with the \texttt{syn()} function of the synthpop R package \citep{gillian2016} using the default parameters. As can be seen in Figure \ref{fig:syndata}, the distributions of the variables in the synthetic data set are very similar to the distributions of the variables in the original data set.

After that, we created an artificial PatientID for each patient and produced rows with varying values on the time-dependent variables. For simplicity, we assumed that if the patient had received adjuvant treatment after surgery, it would always last three years. We created a data point for each year after surgery and predicted whether the recurrence occurred for that year using the Bernoulli model fitted to the original data. We introduced the same time-dependent variables as in \cite{joensuu2014}: Time indicating the year after surgery, AdjOn indicating whether adjuvant treatment was in effect that year, and TimeSinceAdjStopped indicating how many years have passed since adjuvant treatment ended. We created predictions for consecutive years for each patient until cancer recurrence was predicted or 10 years had passed after surgery. In this case, we considered the event time to be censored. The prediction was stored as a binary variable Event. An example set of rows for one patient is shown in Table \ref{table:gist_data_long_example}, and each variable is described in detail in Table \ref{table:variable_descriptions}.

\begin{table}[t]
    \centering
    \small
    \resizebox{\textwidth}{!}{%
        \begin{tabular}{ccccccccccc}
            \textbf{PatientID} & \textbf{Size} & \textbf{AgeAtSurg} & \textbf{MitHPF} & \textbf{GenderMale} & \textbf{Rupture} & \textbf{Gastric} & \textbf{Time} & \textbf{AdjOn} & \textbf{TimeSinceAdjStopped} & \textbf{Event} \\
            \hline
            60 & 75 & 64 & 13 & 0 & 0 & 1 & 1 & 1 & 0 & 0 \\
            60 & 75 & 64 & 13 & 0 & 0 & 1 & 2 & 1 & 0 & 0 \\
            60 & 75 & 64 & 13 & 0 & 0 & 1 & 3 & 1 & 0 & 0 \\
            60 & 75 & 64 & 13 & 0 & 0 & 1 & 4 & 0 & 1 & 0 \\
            60 & 75 & 64 & 13 & 0 & 0 & 1 & 5 & 0 & 2 & 1 \\
        \end{tabular}%
    }
    \caption{Example set of rows for one patient.}
    \label{table:gist_data_long_example}
\end{table}

\begin{table}[t]
    \centering
    \small
    \begin{tabularx}{\textwidth}{@{} l X @{}} 
        \textbf{Variable} & \textbf{Description} \\ 
        \midrule
        PatientID & A unique integer identifier for each patient. \\ \addlinespace
        Size & A continuous numeric variable that represents the size of the tumour at the time of surgery. \\ \addlinespace
        AgeAtSurg & A continuous numeric variable that represents the patient's age at the time of surgery. \\ \addlinespace
        MitHPF & A continuous numeric variable that represents the mitotic count per 50 high power fields at the time of surgery. \\ \addlinespace
        GenderMale & A binary variable that indicates whether the patient is male (1) or female (0). \\ \addlinespace
        Rupture & A binary variable that indicates whether the tumour is ruptured (1) or not (0). \\ \addlinespace
        Gastric & A binary variable that indicates whether the tumour is gastric (1) or in some other location (0). \\ \addlinespace
        Time & A time-dependent integer variable that indicates how many years have passed since surgery. \\ \addlinespace
        AdjOn & A time-dependent binary variable that indicates whether adjuvant treatment is in effect (1) or not (0) in that time interval. \\ \addlinespace
        TimeSinceAdjStopped & A time-dependent integer variable that indicates how many years have passed since adjuvant treatment stopped. \\ \addlinespace
        Event & The binary variable target that indicates whether the recurrence occurs (1) or not (0) in that time interval. 
    \end{tabularx}
    \caption{A detailed description of each variable.}
    \label{table:variable_descriptions}
\end{table}

We also generated a short form of the data set in order to be able to use AFT and PH models in addition to the Bernoulli model. The short form of the data set contains only one row for each patient and does not include time-dependent variables. Instead, adjuvant treatment receival is indicated with a single binary variable AdjTreatm, and the target variable is an integer EventTime, which indicates how many years after surgery the event occurs, as generated using the time-dependent variable model. We now also have a binary variable Censored, which indicates whether the patient is censored. The short form of the data for the patient in Table \ref{table:gist_data_long_example} is shown in Table \ref{table:gist_data_short_example}.

\begin{table}[t]
    \centering
    \renewcommand{\arraystretch}{1.1}
    \small
    \resizebox{\textwidth}{!}{%
        \begin{tabular}{cccccccccc}
            \textbf{PatientID} & \textbf{Size} & \textbf{AgeAtSurg} & \textbf{MitHPF} & \textbf{GenderMale} & \textbf{Rupture} & \textbf{Gastric} & \textbf{AdjTreatm} & \textbf{Censored} & \textbf{EventTime} \\
            \hline
            60 & 75 & 64 & 13 & 0 & 0 & 1 & 1 & 0 & 5 \\
        \end{tabular}
    }
    \caption{The example data point in the short data form.}
    \label{table:gist_data_short_example}
\end{table}

\subsection{Models}

Before fitting any models, the continuous variables Size, AgeAtSurg, and MitHPF were scaled to have 0 mean and 0.5 standard deviation following the recommendation of \cite{gelman2008}. This was done to make the regression coefficients of continuous variables comparable to the coefficients of binary variables. After scaling the variables, we fit four models: the Bernoulli model, the exponential model, the Cox proportional hazards model, and the Weibull accelerated failure time model. The Bernoulli model was fitted to the long data form, and the latter three models were fitted to the short data form. We used weakly informative priors, since we did not have strong prior information available. Detailed description of the models is given below.
\newline

\noindent\textbf{Bernoulli model} \\[0.5em]

\noindent\textit{Model specification}

\begin{align*}
\text{Event}_i &\sim \text{Bernoulli}(p_i) \\
\text{logit}(p_i) &= \beta_0 
+ \beta_1 \cdot \text{AdjOn}_i 
+ \beta_2 \cdot \text{GenderMale}_i 
+ \beta_3 \cdot \text{Rupture}_i 
+ \beta_4 \cdot \text{Gastric}_i \\
&\quad +\ s_1(\text{TimeSinceAdjStopped}_i) 
+ s_2(\text{Time}_i)  \\
 &\quad +\ s_3(\text{Size}_i) 
+ s_4(\text{AgeAtSurg}_i) 
+ s_5(\text{MitHPF}_i)
\end{align*}

\noindent\textit{Prior distributions}

\begin{align*}
\beta_0 &\sim t_3(0, 2.5), \\
\beta_j &\sim \mathcal{N}(0, 2), \quad \text{for all fixed effects } j \\
\beta_k &\sim \mathcal{N}(0, 2), \quad \text{for all spline basis coefficients } k \\
\sigma_l &\sim t_3^+(0, 2.5), \quad \text{for all smoothing standard deviations } l
\end{align*}

\noindent\textbf{Exponential model} \\[0.5em]

\noindent\textit{Model specification}

\begin{align*}
\lambda(t\mid \theta_i) &= \frac{\theta_i \exp(-\theta_i t)}{\exp(-\theta_i t)} = \theta_i \\
\theta_i &= \frac{1}{\mu_i} \\
\log(\mu_i) &= \beta_0 
+ \beta_1 \cdot \text{GenderMale}_i 
+ \beta_2 \cdot \text{Rupture}_i 
+ \beta_3 \cdot \text{Gastric}_i \\
 &\quad +\ s_1(\text{Size}_i) 
+ s_2(\text{AgeAtSurg}_i) 
+ s_3(\text{MitHPF}_i)
\end{align*}

\noindent\textit{Prior distributions}

\begin{align*}
\beta_0 &\sim t_3(2.3, 2.5), \\
\beta_j &\sim \mathcal{N}(0, 2), \quad \text{for all fixed effects } j \\
\beta_k &\sim \mathcal{N}(0, 2), \quad \text{for all spline basis coefficients } k \\
\sigma_l &\sim t_3^+(0, 2.5), \quad \text{for all smoothing standard deviations } l
\end{align*}

\noindent\textbf{Cox PH model} \\[0.5em]

\noindent\textit{Model specification}

\begin{align*}
\lambda_i(t\mid X_i) &= \lambda_0(t)\exp(\eta(X_i)) \\
\eta(X_i) &= \beta_0 
+ \beta_1 \cdot \text{GenderMale}_i 
+ \beta_2 \cdot \text{Rupture}_i 
+ \beta_3 \cdot \text{Gastric}_i \\
 &\quad +\ s_1(\text{Size}_i) 
+ s_2(\text{AgeAtSurg}_i) 
+ s_3(\text{MitHPF}_i)
\end{align*}

\noindent\textit{Prior distributions}

\begin{align*}
\beta_0 &\sim t_3(2.3, 2.5), \\
\beta_j &\sim \mathcal{N}(0, 2), \quad \text{for all fixed effects } j \\
\beta_k &\sim \mathcal{N}(0, 2), \quad \text{for all spline basis coefficients } k \\
\sigma_l &\sim t_3^+(0, 2.5), \quad \text{for all smoothing standard deviations } l
\end{align*}

\noindent
In brms, $\lambda_0(t)$ is modelled using M-splines. The details are omitted here.
\newline

\noindent\textbf{Weibull AFT model} \\[0.5em]

\noindent\textit{Model specification}

\begin{align*}
\lambda_i(t\mid \theta_i, \alpha) &= \theta_i \alpha (\theta_i t)^{\alpha-1}\\
\theta_i &= \frac{\Gamma(1+\frac{1}{\alpha})}{\mu_i} \\
\log(\mu_i) &= \beta_0 
+ \beta_1 \cdot \text{GenderMale}_i 
+ \beta_2 \cdot \text{Rupture}_i 
+ \beta_3 \cdot \text{Gastric}_i \\
 &\quad +\ s_1(\text{Size}_i) 
+ s_2(\text{AgeAtSurg}_i) 
+ s_3(\text{MitHPF}_i)
\end{align*}

\noindent\textit{Prior distributions}

\begin{align*}
\beta_0 &\sim t_3(2.3, 2.5), \\
\beta_j &\sim \mathcal{N}(0, 2), \quad \text{for all fixed effects } j \\
\beta_k &\sim \mathcal{N}(0, 2), \quad \text{for all spline basis coefficients } k \\
\sigma_l &\sim t_3^+(0, 2.5), \quad \text{for all smoothing standard deviations } l \\
\alpha &\sim \Gamma(0.01, 0.01)
\end{align*}

\subsection{Posterior predictive checks}

At the time of writing, the brms implementation of the Cox PH model is very limited. It lacks core functions, such as \texttt{posterior\_predict()} and \texttt{posterior\_epred()}. The implication of this is that posterior predictive checks among many other things cannot be performed for the Cox PH model. Since the implementation of these functions is outside the scope of this paper, we omit the posterior predictive checks for the Cox PH model.

\begin{figure}[t]
  \begin{center}
    \includegraphics[width=0.49\textwidth]{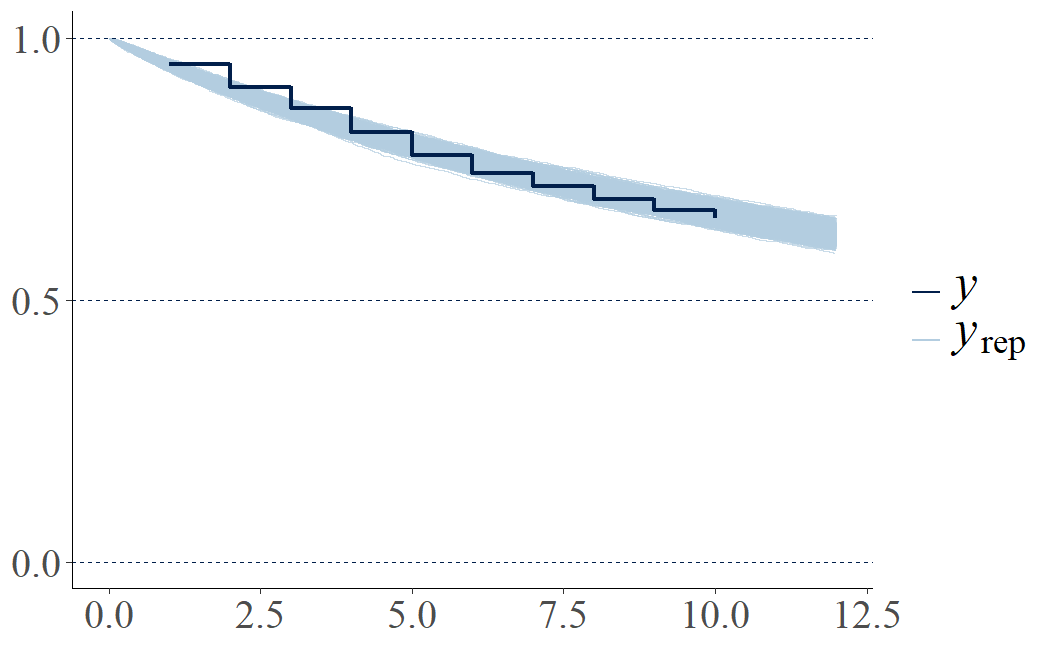}
    \caption{Posterior predictive check for the exponential model.}
    \label{fig:exponential_ppcheck}
  \end{center}
\end{figure}

Since a large portion of the event times in the GIST data set are censored and the censoring process is not modelled, a suitable posterior predictive check for the parametric survival models (the exponential model and the Weibull AFT model) is the Kaplan-Meier overlay plot, where the extrapolation is limited to 20\% of the maximum event or censoring time. The Kaplan-Meier overlay plot for the exponential model is shown in Figure \ref{fig:exponential_ppcheck}. In the simulated data set, EventTime is a discrete variable, so the Kaplan-Meier survival curve for the observations is not very smooth. However, the figure shows that in all discrete time points $1, 2, 3, \dots, 10$, the Kaplan-Meier estimate for the observations falls within the empirical CCDF estimates of the predictive draws. Thus, the predictive model check does not indicate a lack of fit for the exponential model.

\begin{figure}[t]
  \begin{center}
    \includegraphics[width=0.49\textwidth]{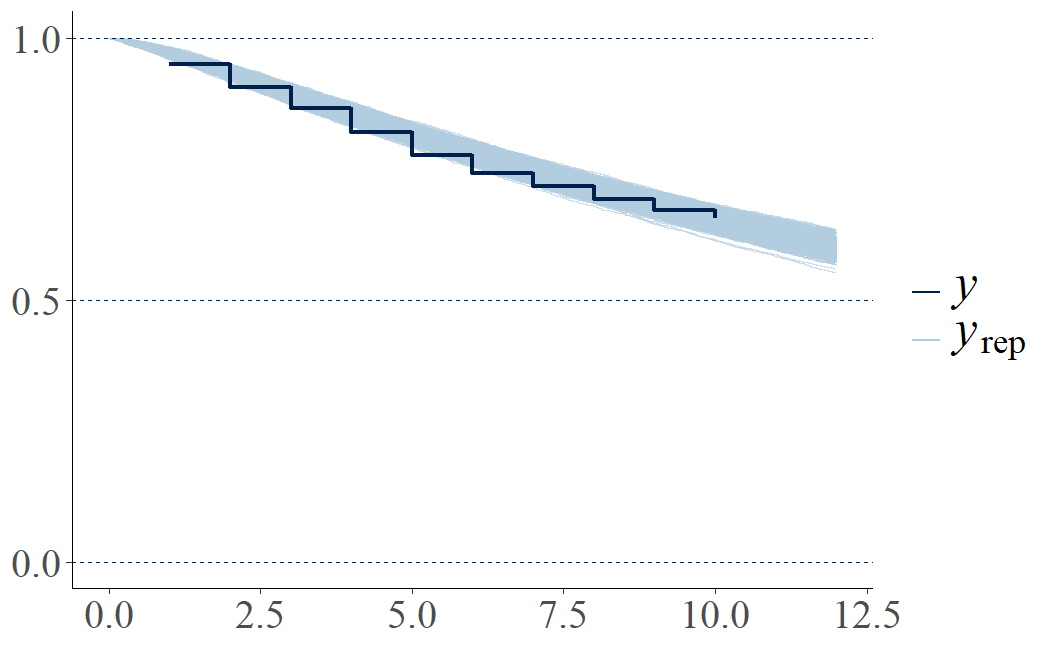}
    \caption{Posterior predictive check for the Weibull AFT model.}
    \label{fig:weibull_ppcheck}
  \end{center}
\end{figure}

The posterior predictive check for the Weibull AFT model is shown in Figure \ref{fig:weibull_ppcheck}. The figure shows that in many discrete time points, the Kaplan-Meier estimate for the observations falls below the empirical CCDF estimates of the predictive draws. This indicates a lack of fit for the Weibull AFT model.

\begin{figure}[t]
  \begin{center}
    \includegraphics[width=0.65\textwidth]{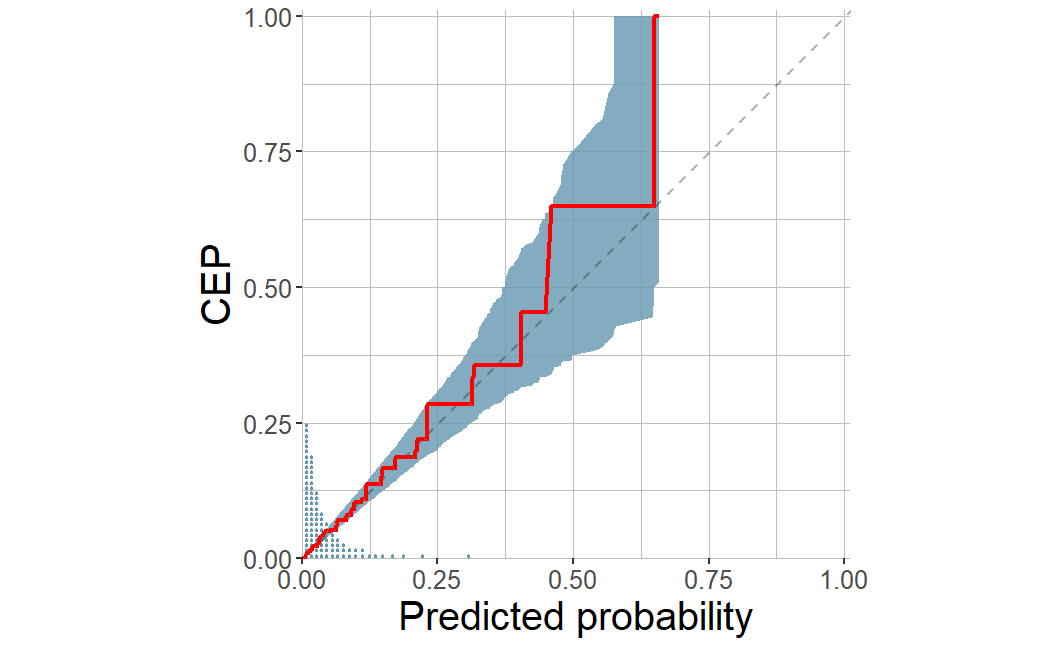}
    \caption{Posterior predictive check for the Bernoulli model.}
    \label{fig:bernoulli_ppcheck}
  \end{center}
\end{figure}

As the Bernoulli model has a binary outcome, the PAV-adjusted calibration plot is the recommended posterior predictive check for it. This is visualised in Figure \ref{fig:bernoulli_ppcheck}. The calibration curve does not exceed the consistency bands, so there is no apparent lack of fit.

\begin{figure}[t]
  \begin{center}
    \includegraphics[width=0.65\textwidth]{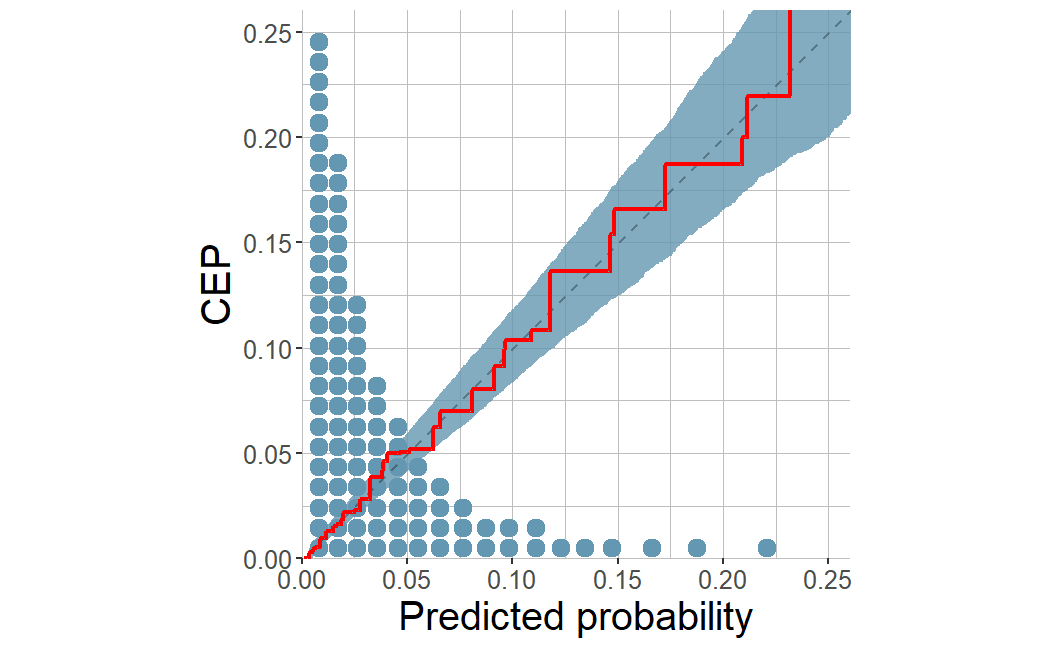}
    \caption{Zoomed posterior predictive check for the Bernoulli model.}
    \label{fig:bernoulli_ppcheck_zoomed}
  \end{center}
\end{figure}

In the long format of the GIST data set, most data points do not have the event occur in them. Therefore, the predicted probability of cancer recurrence is near zero for most data points. This is indicated in Figure \ref{fig:bernoulli_ppcheck} by the small dots that mostly fall in the range $[0.00, 0.25]$. In this kind of case, it is useful to zoom in on the most dense region, as is done in Figure \ref{fig:bernoulli_ppcheck_zoomed}. Even the zoomed figure does not show the calibration curve exceeding the confidence band, so there does not seem to be a lack of fit for the Bernoulli model.

\subsection{Predictive Model Comparison}

When fitted with brms, the exponential model, the Cox PH model, and the Weibull AFT model have the same target variable. This means that the \texttt{loo\_compare()} function in the loo R package does not give any warning when we use it to perform the predictive comparison, as long as the time scales of the target variables are the same. The result of \texttt{loo\_compare()} is presented in Table \ref{table:regular_psis_loo_cv}.

\begin{table}[H]
\begin{center}
\renewcommand{\arraystretch}{1.3}
\begin{tabular}
{p{0.25\textwidth}p{0.25\textwidth}p{0.25\textwidth}}
Model & $\Delta\widehat{\text{elpd}}_{\text{LOO}}$ &  $\text{se}(\Delta\widehat{\text{elpd}}_{\text{LOO}})$\\
    \hline
Cox PH & - & - \\
Weibull AFT & -117.5 & 9.8 \\
Exponential & -173.4 & 6.6 \\
\end{tabular}
\end{center}
\caption{Result of model comparison with regular PSIS-LOO CV.}
\label{table:regular_psis_loo_cv}
\end{table}

The table indicates that the Cox PH model is superior to the Weibull AFT model and the exponential model. However, it should be noted that a significant portion of event times in the GIST data are censored. In this kind of case, pointwise elpds are a mixture of probabilities and densities, as stated in Subsection \ref{subsec:predictive_comparison_ph_aft_censoring}.

In order to avoid mixing probabilities and densities, we can change the prediction task so that we only deal with probabilities. The simplest way to do this is to move to predicting the probability that an event has happened before some fixed time point, and perform PSIS-LOO CV using these predictions and their log scores. Currently, this is not possible with the Cox PH model due to the limitations of brms, but it is possible with the Weibull AFT model and the exponential model.

Since the Weibull AFT model and the exponential model are parametric survival models, we can use their cumulative distribution functions to compute the probability that an event happens before some fixed time point. For positive time points $t$, the CDF of the Weibull distribution is
\begin{equation}
    F(t)=1-\exp\left(-\left(\theta t\right)^\alpha\right),
\end{equation}
where $\alpha$ is the shape parameter and $\theta$ is the rate parameter. In brms, the model is parameterised in terms of the mean $\mu$ instead of $\theta$, but they are related by the equation $\theta = \frac{\Gamma(1+\frac{1}{\alpha})}{\mu}$. On the other hand, the CDF of the exponential distribution is 
\begin{equation}
    F(t)=1-\exp(-\theta t),
\end{equation}
where $\theta$ is the rate parameter. Again, brms parameterisation uses the mean $\mu$ instead of $\theta$, but these two parameters are related by equation $\theta=\frac{1}{\mu}$.

We predicted the probability that the event occurs at most 5 years after surgery, computed log scores for the predicted probabilities, and performed a PSIS-LOO CV to compare the predictive performance of the Weibull AFT model and the exponential model. The result is presented in Table \ref{table:5_years_psis_loo_cv}.

\begin{table}[H]
\begin{center}
\renewcommand{\arraystretch}{1.3}
\begin{tabular}
{p{0.25\textwidth}p{0.25\textwidth}p{0.25\textwidth}}
Model & $\Delta\widehat{\text{elpd}}_{\text{LOO}}$ &  $\text{se}(\Delta\widehat{\text{elpd}}_{\text{LOO}})$\\
    \hline
Exponential & - & - \\
Weibull AFT & -17.1 & 4.4 \\
\end{tabular}
\end{center}
\caption{Result of PSIS-LOO CV model comparison in a dichotomised prediction task.}
\label{table:5_years_psis_loo_cv}
\end{table}

The table shows that the exponential model outperforms the Weibull AFT model in this prediction task. It should be noted that, while the dichotomisation simplifies the prediction task, it loses quite a bit of information. As explained in Subsection \ref{subsec:predictive_comparison_ph_aft_censoring}, it is also possible to perform a comparison that loses less information while still comparing predictive probabilities. This can be done by discretising time to some fixed length intervals and predicting the probability that an event occurs in each of the intervals. However, this approach is more complicated to implement.

Both the simple dichotomisation approach and the more complicated time discretisation approach allow the inclusion of the Bernoulli model in the comparison. Cross-validation for the Bernoulli model requires leaving out all data points corresponding to one patient and PSIS may fail as the posterior of a flexible model can change too much for importance weighting based approach to work reliably. Performing cross-validation for the Bernoulli model may require fitting the model separately for each patient, which can be computationally expensive.

\begin{figure}[p]
    \centering % Center the entire figure content
    \begin{subfigure}{0.49\textwidth}
        \centering
        \includegraphics[width=\linewidth]{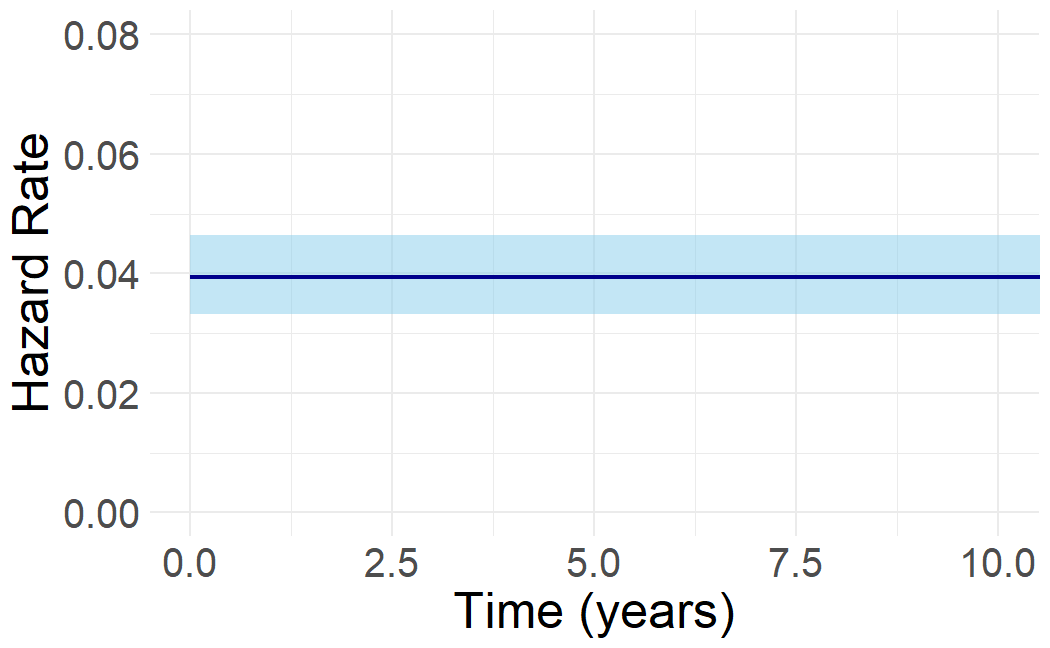}
        \caption{Adjuvant treatment given.}
        \label{fig:a_exponential_hazard_treat}
    \end{subfigure}
    \hfill
    \begin{subfigure}{0.49\textwidth}
        \centering
        \includegraphics[width=\linewidth]{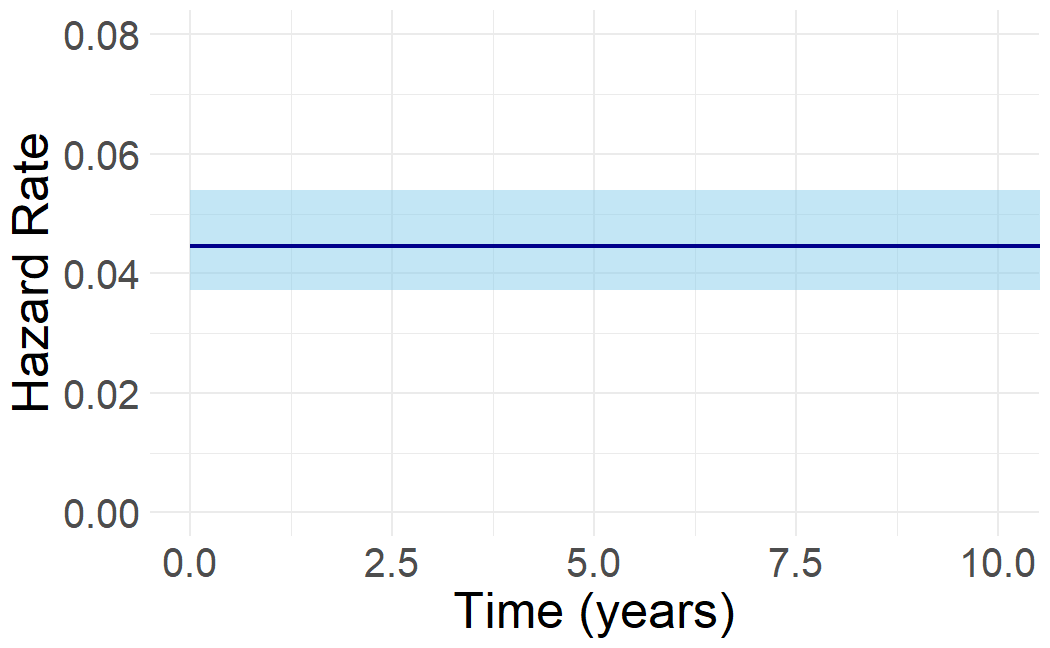}
        \caption{No adjuvant treatment given.}
        \label{fig:b_exponential_hazard_no_treat}
    \end{subfigure}
    % The main figure caption
    \caption{Hazard for the example patient with and without treatment as predicted with the exponential model. Dark blue lines illustrate the median and light blue ribbons illustrate the 95\% credible interval. The hazard of the exponential model stays constant across time. It is slightly lower in the case where adjuvant treatment is administered than in the case where it is not administered.}
    \label{fig:exponential_hazard}
\end{figure}

\begin{figure}[p]
    \centering
    \begin{subfigure}{0.49\textwidth}
        \centering
        \includegraphics[width=\linewidth]{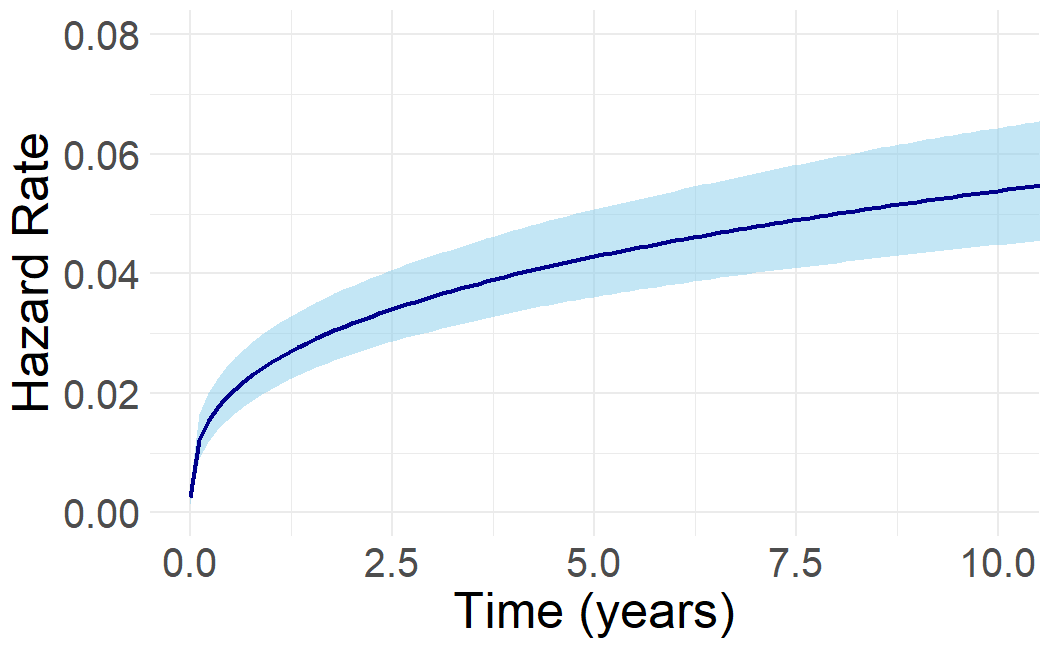}
        \caption{Adjuvant treatment given.}
        \label{fig:a_weibull_hazard_treat}
    \end{subfigure}
    \hfill
    \begin{subfigure}{0.49\textwidth}
        \centering
        \includegraphics[width=\linewidth]{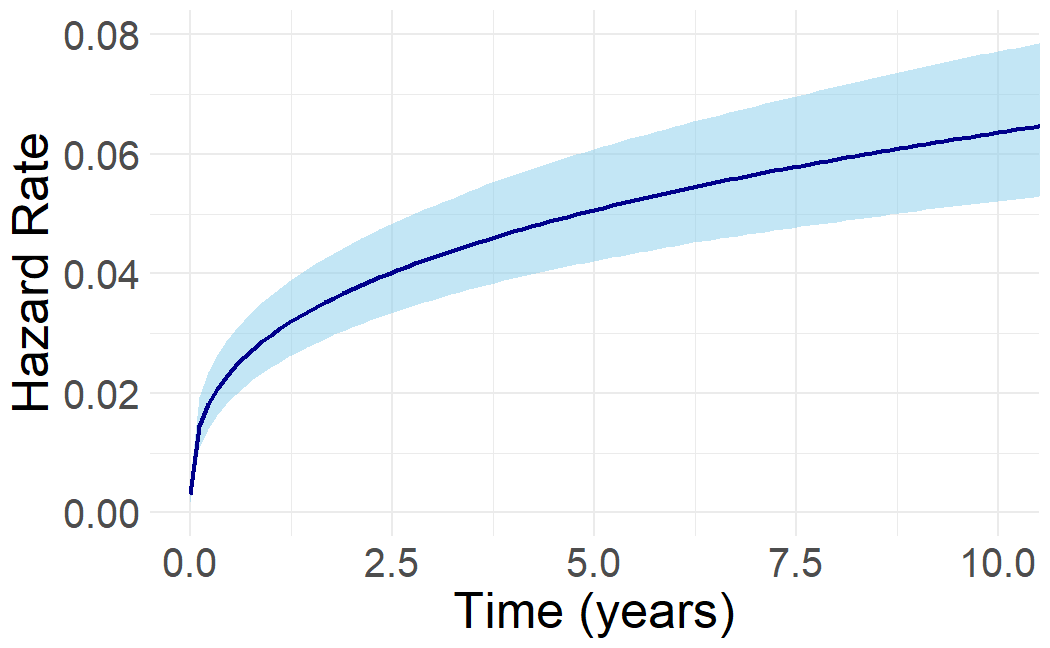}
        \caption{No adjuvant treatment given.}
        \label{fig:b_weibull_hazard_no_treat}
    \end{subfigure}
    \caption{Hazard for the example patient with and without treatment as predicted with the Weibull AFT model. Dark blue lines illustrate medians and light blue ribbons illustrate 95\% credible intervals. The hazard rate of the Weibull AFT model increases over time. In all time points, the median of the hazard is slightly lower in the case where adjuvant treatment is administered than in the case where it is not administered.}
    \label{fig:weibull_hazard}
\end{figure}

\begin{figure}[p]
    \centering
    \begin{subfigure}{0.49\textwidth}
        \centering
        \includegraphics[width=\linewidth]{figures/bernoulli_recurrence_probability_treat.png}
        \caption{Adjuvant treatment given.}
        \label{fig:a_bernoulli_recurrence_probability_treat}
    \end{subfigure}
    \hfill
    \begin{subfigure}{0.49\textwidth}
        \centering
        \includegraphics[width=\linewidth]{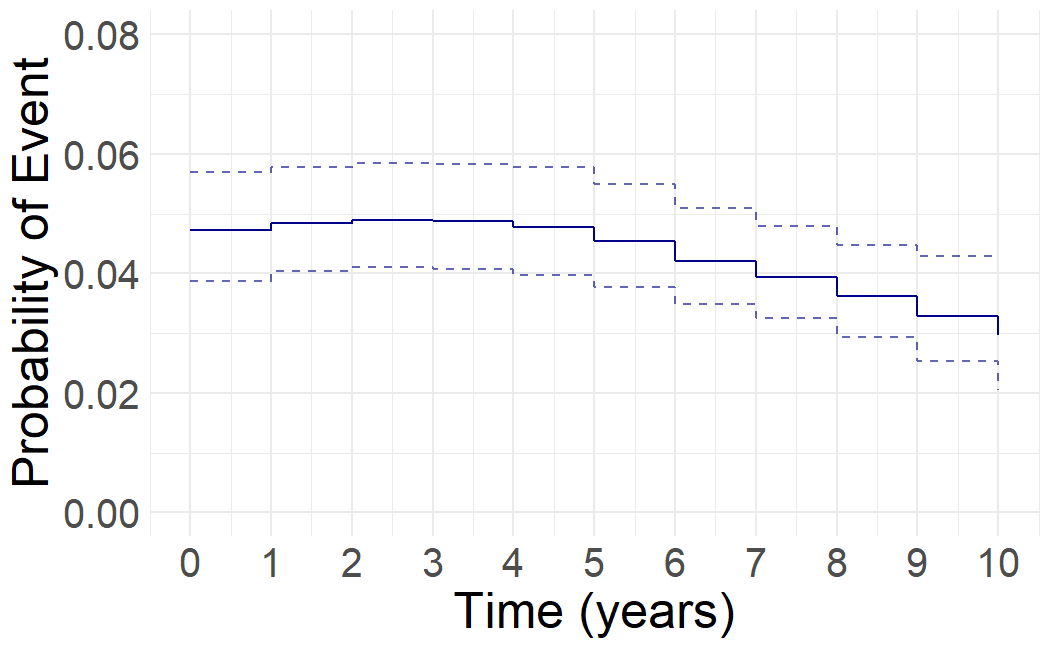}
        \caption{No adjuvant treatment given.}
        \label{fig:b_bernoulli_recurrence_probability_no_treat}
    \end{subfigure}
    \caption{Recurrence probabilities for the example patient with and without treatment as predicted with the Bernoulli model. Solid lines illustrate medians and dashed lines illustrate the bounds of 95\% credible intervals. Figure \ref{fig:a_bernoulli_recurrence_probability_treat} shows the case where adjuvant treatment is administered for the patient for three years. When the adjuvant treatment is in effect, the probability of recurrence remains low. When the treatment ends, the probability jumps higher and starts decreasing over time. On the other hand, Figure \ref{fig:b_bernoulli_recurrence_probability_no_treat} shows the counterfactual case where adjuvant treatment is not administered. In this case, the probability of recurrence is high at the beginning and starts decreasing after a while.}
    \label{fig:bernoulli_recurrence_probability}
\end{figure}

That being said, in the GIST data, adjuvant treatment has a clear time-dependent effect on the probability of recurrence, which allows determining the superiority of the Bernoulli model without any computational comparison. This can be demonstrated visually. In Figure \ref{fig:exponential_hazard} and Figure \ref{fig:weibull_hazard}, we visualise the predicted hazard rate from the exponential model and the Weibull AFT model, respectively. Similarly, Figure \ref{fig:bernoulli_recurrence_probability} illustrates the predicted recurrence probability from the Bernoulli model for each time interval. This can be thought of as a discrete time equivalent of the hazard rate.

The prediction is done for the example patient, whose values are shown in Table \ref{table:gist_data_long_example} and Table \ref{table:gist_data_short_example}. The prediction is performed for a case where the patient is treated and for a counterfactual case where the patient is not treated. This clearly demonstrates how different models handle the effect of the treatment.

As can be seen in the figures, the Bernoulli model handles the time-dependent effect of adjuvant treatment on the recurrence probability in a way that is not possible in either the exponential model or the Weibull AFT model due to their model assumptions and inability to include variables with time-dependent effects. Since brms has a very limited implementation of the Cox PH model at the moment, we cannot visualise the predicted hazard from the Cox PH model. However, the proportional hazards assumption makes the Cox PH model incapable of capturing the time-dependent effect of the treatment visualised in Figure \ref{fig:bernoulli_recurrence_probability}. Therefore, we can safely state that the Bernoulli model is superior to all other models presented in this paper when it comes to predicting the probability of recurrence of a gastrointestinal stromal tumour.

\section{Recommendations}
\label{sec:recommendations}

In this section, we summarise our recommendations for predictive model checking and predictive model comparison for survival models. This section only contains the recommendations for different scenarios without justifications. Justifications and examples are included in the earlier sections of this paper.

\subsection{Predictive model checking}

\noindent
\textbf{Scenario 1: Parametric survival models in the absence of censored event times} \indent If the data set does not contain censored event times, many diagnostic plots, such as the intervals plot, the PIT-ECDF plot, and the Kaplan-Meier overlay plot, are applicable for assessing predictions of parametric survival models. If some observations are left-truncated, we recommend taking the truncation into account when plotting the survival curves in the Kaplan-Meier overlay plot.
\newline

\noindent
\textbf{Scenario 2: Parametric survival models in the presence of censored event times} \indent In case there are censored event times and the censoring process is not modelled, the intervals plot and the PIT-ECDF plot become unusable, at least without the imputation of censored event times. The Kaplan-Meier overlay plot remains usable, but we recommend fine-tuning it by cutting the x-axis so that the plot extrapolates no more than 20\% beyond the furthest observation. In addition, we recommend imputing censored event times and plotting a survival curve that includes them with a different colour than the survival curve for just the observed data points. The recommendation of taking left-truncation into account in the survival curves still applies.
\newline

\noindent
\textbf{Scenario 3: The Bernoulli model} \indent We recommend following the recommendation by \cite{sailynoja2025} and utilising the PAV-adjusted calibration plots. We extend their recommendation by suggesting to examine a zoomed plot in addition to the original one in case a large portion of predicted probabilities fall on a specific interval.

\subsection{Predictive model comparison}

\noindent
\textbf{Scenario 1: PH and AFT models in the absence of censored event times} \indent In the case of comparing only the PH and AFT models in the absence of censored event times, one can apply PSIS-LOO CV with the usual recommendations, as long as one makes sure that the event times are on the same time scale in all models.
\newline

\noindent
\textbf{Scenario 2: PH and AFT models in the presence of censored event times} \indent In the presence of censored event times, there is the possibility that the chosen time scale causes problems in model comparison. If one does not wish to carefully inspect that the time scale does not cause problems, one can discretise time, and hence do the model comparison with probabilities.
\newline

\noindent
\textbf{Scenario 3: PH, AFT, and Bernoulli models} \indent When the Bernoulli model is included in the comparison, the computational comparison is recommended to be performed with probabilities, regardless of whether there are censored event times present. Furthermore, the cross-validation for the Bernoulli model has to be performed so that all the rows corresponding to the selected subject are left out. However, in case the data set contains variables with clear time-dependent effects, the superiority of the Bernoulli model can often be determined without any computational comparison.
\newline

\section{Discussion}
\label{sec:discussion}

\subsection{Limitations}

The paper has some limitations on the extent to which some of the topics are covered. For example, we only included models that were possible to fit with brms and left out models that were not. Also, we only focused on the log score when we covered model comparison, even though there are many other options for scoring metrics.

An important limitation regarding the case study of gastrointestinal stromal tumours is the fact that we had to use simulated data instead of real data. We also left out some covariates of the original data set that were not used in previous GIST papers. Again, we want to emphasise that the case study should be interpreted as merely a demonstration of the methods, and the exact results do not hold much value.

\subsection{Future Directions}

From the limitations of this paper we can naturally move on to potential future research directions. First of all, the research could be extended to survival models outside the brms framework. Secondly, in addition to the log scores, model comparison could focus on other scoring rules that directly address the issue with censored event times. These kinds of scoring rules include Survival-CRPS \citep{avati2019} and strictly proper censoring-adjusted separable scoring rule \citep{alberge2024}. Third, even though we implemented some of our findings in software tools, some remain unimplemented. Our recommendation to impute censored event times, when performing predictive model checks could be implemented to brms and bayesplot in the future. Software implementations could also be made to make it easier to compare models using our recommended method based on changing the prediction task so that all pointwise elpds are probabilities rather than densities. Fourth, as noticed while writing the paper, the Cox PH model does not have a predictive model check. This could be created and implemented. Finally, the methodology presented in this paper could be applied in the analysis of real data on gastrointestinal stromal tumours. 

\subsection*{Acknowledgements}
We thank Heikki Joensuu and Joris van Sabben for letting us collaborate on their project on gastrointestinal stromal tumour.
\bibliography{library}
\end{document}